\documentclass[
 superscriptaddress,
 longbibliography,
 nostarredaffiliations,
 twocolumn,
 amsmath,amssymb,
]{revtex4-2}
\usepackage[utf8]{inputenc} 
\usepackage[T1]{fontenc}
\usepackage{booktabs} 
\usepackage{lmodern} 
\usepackage{graphicx}
\usepackage{dcolumn}
\usepackage{bm}
\usepackage[colorlinks=true,linkcolor=blue,urlcolor=blue,citecolor=blue]{hyperref}

\usepackage{color}
\usepackage{xcolor}
\usepackage{siunitx}
\usepackage{times}
\usepackage{braket}
\usepackage{siunitx}
\usepackage{natbib} 
\usepackage{tabularx}
\usepackage{mathtools}
\usepackage{rotating}
\usepackage{tikz}
\usepackage{newfloat}

\DeclareFloatingEnvironment[
fileext=lof, 
listname={List of Extended Data Figures}, 
name={Extended Data Fig}, 
placement=htp 
]{extfig}

\definecolor{singletblue}{HTML}{005DFF}
\definecolor{singletred}{HTML}{FF0027}

\makeatletter
\def\@hangfrom@section#1#2#3{\@hangfrom{#1#2}#3}
\def\@hangfroms@section#1#2{#1#2}
\makeatother

\allowdisplaybreaks

\newcommand{\nhat}{\hat{n}}

\newcommand{\nhole}{\hat{n}^{h}}

\newcommand{\Sx}{\hat{S}^{x}}
\newcommand{\Sz}{\hat{S}^{z}}

\newcommand{\Rb}{\ensuremath{^{87}\text{Rb}~}}

\newcommand{\degree}{\ensuremath{^{\circ}}}

\newcommand{\tJ}{$t$-$J$ }

\begin{document}

\title{Kinetically-induced bound states in a frustrated Rydberg tweezer array}

\author{Mu~Qiao$^{\ddagger}$}
\altaffiliation{These authors contributed equally to this work.}
\affiliation{
Universit\'{e} Paris-Saclay, Institut d'Optique Graduate School, CNRS, Laboratoire Charles Fabry, 91127 Palaiseau Cedex, France
}
\email{mu.q.phys@gmail.com}

\author{Romain~Martin}
\altaffiliation{These authors contributed equally to this work.}
\affiliation{
Universit\'{e} Paris-Saclay, Institut d'Optique Graduate School, CNRS, Laboratoire Charles Fabry, 91127 Palaiseau Cedex, France
}

\author{Lukas~Homeier}
\altaffiliation{These authors contributed equally to this work.}
\affiliation{JILA and Department of Physics, University of Colorado, Boulder, CO, 80309, USA}
\affiliation{Center for Theory of Quantum Matter, University of Colorado, Boulder, CO, 80309, USA}

\author{Ivan~Morera}
\altaffiliation{These authors contributed equally to this work.}
\affiliation{Institute for Theoretical Physics, ETH Zurich, 8093 Zurich, Switzerland.}

\author{Bastien~G\'ely}
\affiliation{
Universit\'{e} Paris-Saclay, Institut d'Optique Graduate School, CNRS, Laboratoire Charles Fabry, 91127 Palaiseau Cedex, France
}

\author{Lukas~Klein}
\affiliation{
Universit\'{e} Paris-Saclay, Institut d'Optique Graduate School, CNRS, Laboratoire Charles Fabry, 91127 Palaiseau Cedex, France
}

\author{Yuki~Torii~Chew}
\affiliation{
Universit\'{e} Paris-Saclay, Institut d'Optique Graduate School, CNRS, Laboratoire Charles Fabry, 91127 Palaiseau Cedex, France
}

\author{Daniel~Barredo}
\affiliation{
Universit\'{e} Paris-Saclay, Institut d'Optique Graduate School, CNRS, Laboratoire Charles Fabry, 91127 Palaiseau Cedex, France
}
\affiliation{
Nanomaterials and Nanotechnology Research Center (CINN-CSIC), Universidad de Oviedo (UO), Principado de Asturias, 33940 El Entrego, Spain
}

\author{Thierry~Lahaye}
\affiliation{
Universit\'{e} Paris-Saclay, Institut d'Optique Graduate School, CNRS, Laboratoire Charles Fabry, 91127 Palaiseau Cedex, France
}

\author{Eugene~Demler}
\affiliation{Institute for Theoretical Physics, ETH Zurich, 8093 Zurich, Switzerland.}

\author{Antoine~Browaeys}
\email{antoine.browaeys@institutoptique.fr}
\affiliation{
Universit\'{e} Paris-Saclay, Institut d'Optique Graduate School, CNRS, Laboratoire Charles Fabry, 91127 Palaiseau Cedex, France
}

\date{\today}
\begin{abstract}
Understanding how particles bind into composite objects is a ubiquitous theme in physics, from the formation of molecules~\cite{ChemicalBond1962} to hadrons in quantum chromodynamics~\cite{hadron2018} and the pairing of charge carriers in superconductors~\cite{BCS1957}. The formation of bound states usually originates from attractive interactions between particles. However, the binding can also arise purely from the motion of dopants due to kinetic frustration~\cite{Batista2017,Attraction2024}, which is potentially related to unconventional pairing in moir\'e materials~\cite{Batista2017,Attraction2024,magnonic_sc}. Here, we report the first direct observation of kinetically-induced bound states between holes and magnons using a Rydberg atom array quantum simulator of the bosonic \tJ model in frustrated ladders and 2D lattices~\cite{Homeier2024,tJ_qiao2025}. First, we demonstrate the formation of mobile one-hole-one-magnon bound states. We then construct three-particle one-hole-two-magnon bound states and reveal the underlying binding mechanism by observing kinetically-induced singlet correlations. Finally, we investigate how mobile dopants structure their magnetic environment in a spin-balanced 2D triangular lattice, showing that a hole induces $120\degree$ antiferromagnetic order, while a doublon dopant generates in-plane ferromagnetic correlations. Our results demonstrates compelling evidence of kinetically-induced binding, opening a new avenue to understand novel pairing mechanisms in correlated quantum materials like superconductors in moir\'e superlattices.

\end{abstract}

\maketitle

Conventionally, bound states are formed due to attractive interactions between particles. While having particles close to each other increases their kinetic energy, lowering of the interaction energy dominates, resulting in a state of total negative energy. A different mechanism of particle binding can be found in p-wave pairing in superfluid $^3$He~\cite{Helium1997} and d-wave electron pairing in high-Tc cuprates~\cite{Dopping2006,ScalapinoSC2012}. In these systems, the dominant interaction between fermions is repulsive, yet interaction energy can be lowered by forming pairs with non-zero angular momentum. Theoretical proof of the possibility of pairing in systems with repulsive interactions has been provided by the Kohn-Luttinger theorem~\cite{KohnLuttinger1965}. More recently, a novel binding mechanism based on kinetic frustration has been proposed. In this scenario, bringing particles together alleviates the frustration in their individual propagation paths and lowers kinetic energy~$\propto t$ of the system \cite{Batista2017,Attraction2024}. This kinetically-induced binding is at the origin of the magnetic interactions in systems where the usual superexchange interactions~$\propto J$ are too weak, such as Wigner crystals of electrons in two-dimensions (2D) \cite{Interstitial2022} and moir\'e superlattices in transition metal dichalcogenides (TMDCs) \cite{TMD_Tang2020,TMD_Ciorciaro2023}.

Kinetic magnetism has been observed across diverse platforms, including moiré heterostructures~\cite{TMD_Ciorciaro2023,TMD_MakKinFai2024}, ultracold atoms in optical lattices~\cite{OL_Xu2023,OL_Lebrat2024,OL_Prichard2024,OL_Prichard2025}, and small arrays of quantum dots~\cite{QD_Dehollain2020}. The observed magnetic order occurs at temperatures exceeding the small magnetic energy scales of conventional antiferromagnetic superexchange interactions in these systems, pointing instead to a kinetic origin. In non-frustrated lattices, such as the square lattice, constructive interference between different propagation paths of charge carriers in a polarized background reduces their kinetic energy, leading to Nagaoka-Thouless ferromagnetism~\cite{Nagaoka1966,Thouless1965}. 
By contrast, non-bipartite lattices such as the triangular lattice, can exhibit kinetic frustration: itinerant doped holes moving in polarized spin backgrounds experience destructive quantum interference between their different propagation paths. To alleviate this frustration, antiferromagnetic spin correlations emerge~\cite{HS2005}. The same mechanism underlies the formation of kinetically-induced bound states when starting from a strongly polarized state. When starting from a strongly polarized state, an itinerant charge carrier can alleviate kinetic frustration by binding to a nearby spin excitation (a magnon), making it energetically favorable for the two particles to propagate together as a bound state~\cite{Batista2017,High2023,Exploring2024,Attraction2024}. Larger multi-body composites involving multiple magnons and holes have also been theoretically predicted~\cite{Batista2017,Attraction2024}. Their self-organization leads to emergent many-body phases beyond standard mean-field descriptions, including unconventional paired phases, magnetization plateaus and phase separation~\cite{Attraction2024,High2023,Plateaus2023,Itinerant2024}. Despite the rich physics resulting from kinetically-induced binding, a direct microscopic experimental observation of this exotic mechanism remains elusive.

\begin{figure*}
\mbox{}
\includegraphics[width=\textwidth]{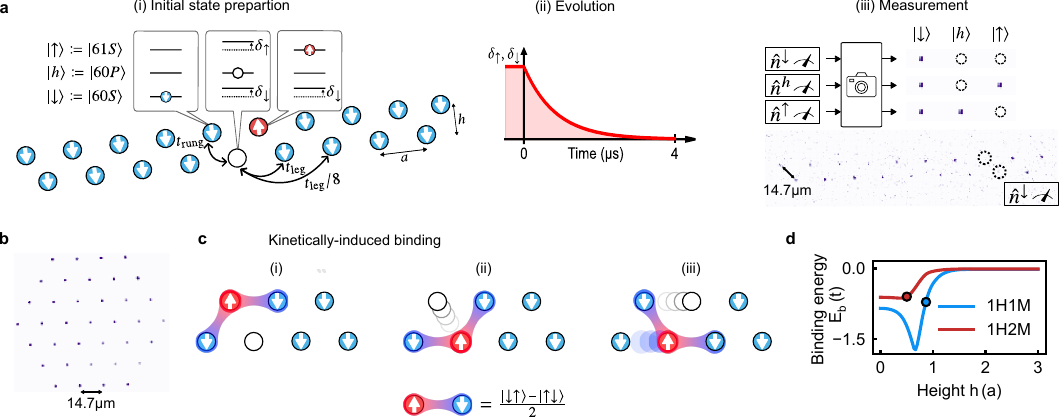}
\caption{\label{fig:setup} 
{\bf Experimental setup and the mechanism of kinetically-induced binding.}
\textbf{a,} Experimental setup and sequence. The spin-down $\ket{\downarrow}$, hole $\ket{h}$, and spin-up $\ket{\uparrow}$ states are mapped onto the $\ket{60S_{1/2}, m_{J}=1/2}$, $\ket{60P_{3/2}, m_{J}=1/2}$, and $\ket{61S_{1/2}, m_{J}=1/2}$  Rydberg levels of $^{87}$Rb atoms. The atoms are arranged in a triangular ladder geometry, characterized by intra-leg spacing $a$ and inter-leg spacing $h$. The sequence consists of (i) initial state preparation using local light shifts, (ii) time evolution during which we ramp down the light shift, and (iii) measurement of final state occupations. The populations in the spin-down ($\hat{n}^{\downarrow}$), hole ($\hat{n}^{h}$), and spin-up ($\hat{n}^{\uparrow}$) bases are measured in separate experimental runs, where the state at each site is mapped to the presence or absence of an atom. The image below shows a single shot from a spin-down measurement capturing a bound pair.
\textbf{b,} Fluorescence image of a 37-site 2D triangular lattice.
\textbf{c,} Mechanism of kinetically-induced binding. (i) A magnon forms a singlet state with adjacent spins surrounding a hole, creating a bound state. (ii) The hole moves to magnon's position, an internal dynamic of the bound state that remains kinetically unfrustrated. (iii) The hole hops to a neighboring spin-down site, effectively moving the bound state by one lattice site.
\textbf{d,} Theoretical calculation of the binding energy ($E_b/t$) versus the ladder's aspect ratio ($h/a$) for a single hole bound to one magnon (1H1M, blue) and two magnons (1H2M, red). The dot marks a set of parameters used in the experiments.
}
\end{figure*}

Here, we report the first direct observation of kinetically-induced bound states between holes and magnons using a bosonic \tJ quantum simulator implemented with Rydberg atom arrays~\cite{tJ_qiao2025,Homeier2024,Annabelle_2024}. While early work on cold atoms, motivated by cuprates, focused on magnetic polarons in spin-balanced mixtures, our study is instead inspired by recent experiments in moir\'e systems that observed both kinetic magnetism and superconductivity arising next to correlated insulating states~\cite{TMD_Tang2020,TMD_Tang2023,TMD_MakKinFai2024,SC_MATBG_Cao2018,SC_TDBG_Liu2020,SC_TMD_Guo2025,SC_TMD_Xia2025}. We realize these systems in both triangular ladders and 2D triangular lattices, providing the first microscopic characterization of these composite particles. 

Our results are threefold. First, we present the direct observation of one-hole-one-magnon (1H1M) bound states, characterizing these composite quasiparticles and their delocalization in both triangular ladders and 2D arrays. Second, we construct and characterize one-hole-two-magnon (1H2M) three-body bound states on the ladder, then measure the local antiferromagnetic spin correlations induced by the hole’s motion to provide microscopic insight into the binding mechanism. Third, we investigate how mobile dopants structure their magnetic environment in a spin-balanced 2D triangular lattice, showing that a hole dopant induces local antiferromagnetic order, while a doublon dopant generates a transverse ferromagnetic spin bag.

\section*{Experimental system}
\label{sec:experimental_system}

\begin{figure*}
\mbox{}
\includegraphics[width=\textwidth]{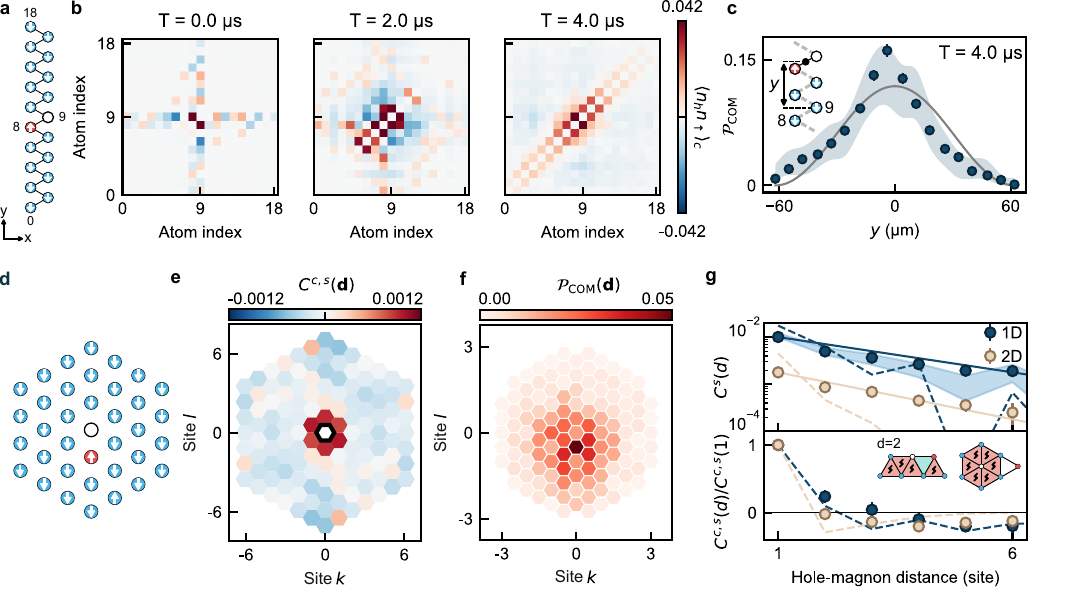}
\caption{\label{fig:1H1M_bound_state} 
{\bf Kinetically-induced one-hole-one-magnon bound state in ladders.}
\textbf{a,} Initial state on a 19-site triangular ladder, with a hole (white) and magnon (red) on adjacent sites.
\textbf{b,} Time evolution of the hole-magnon connected correlations, $\langle n_i^h n_j^\uparrow \rangle^{s}_{c}$.
\textbf{c,} Cut of the COM probability distribution, $\mathcal{P}_{\text{COM}}$, along the center of the ladder (see inset). Experimental data (blue circles) is compared to the corresponding $\sin^2$ distribution (solid grey line). The shaded area represents numerically simulated distribution including errors.
\textbf{d,} Initial state on a 37-site 2D triangular lattice.
\textbf{e,} Hole-magnon connected and symmetrized correlations in the hole's reference frame, $C^{c, s}(\mathbf{d})$, at time $T=5\,\mu\text{s}$.
\textbf{f,} The delocalized COM probability distribution, $\mathcal{P}_{\text{COM}}$, at $T=5\,\mu\text{s}$ of the 1H1M bound pair.
\textbf{g,} Top: Hole-magnon non-connected and symmetrized correlations $C^s(d)$ versus distance for ladders~(blue circles) and 2D arrays~(tan circles). Solid lines represent exponential fit of the correlation. Shaded areas are the numerical simulation including errors. Bottom: Normalized connected correlations $C^{c,s}(d)/C^{c,s}(1)$. Dashed lines correspond to the theoretical ground state. The inset shows the dominant kinetic contributions of the binding, and the red plaquettes are kinetically frustrated while the green one's frustration is alleviated. Error bars denote one standard error estimated via bootstrap, and are smaller than marker size.
}
\end{figure*}

Our experimental setup relies on arrays of individual $^{87}$Rb atoms trapped in optical tweezers. We implement a hard-core bosonic $t$--$J$ model by encoding spin up, spin down, and hole states into three distinct Rydberg levels (see Methods): $\ket{\downarrow} \coloneqq \ket{60S_{1/2}, m_{J} = 1/2}$, $\ket{h} \coloneqq \ket{60P_{3/2}, m_{J} = 1/2}$, $\ket{\uparrow} \coloneqq \ket{61S_{1/2}, m_{J} = 1/2}$. Dipole-dipole interactions between them realize a kinetically-frustrated bosonic $t$--$J$ Hamiltonian $\hat{H}_{tJ} = \hat{H}_{t}+\hat{H}_{J}$~\cite{Homeier2024, tJ_qiao2025} with

\begin{equation}
\label{eq:tJ_Hamiltonian}
\begin{aligned}
&\hat{H}_{t} =+\hbar\sum_{i<j}\sum_{\sigma=\downarrow,\uparrow
}t_{\sigma}\frac{a^3}{{r}_{ij}^3}\hat{\mathcal{P}}_{G}\left(\hat{b}_{i,\sigma}^\dagger\hat{b}_{j,\sigma}+\text { h.c. }\right)\hat{\mathcal{P}}_{G}, \\
&\hat{H}_{J} = \hbar\sum_{i<j}\frac{a^6}{{r}_{ij}^6}\left[J_{z}\hat{S}_i^z \hat{S}_j^z+\frac{J_{\perp}}{2}\left(\hat{S}_i^{+} \hat{S}_j^{-}+\text {h.c. }\right)\right]\ , 
\end{aligned}
\end{equation}
where $\hat{b}^{\dagger}_{i,\sigma}$ is the creation operator for a hard-core boson with spin $\sigma$ at site $i$, $\hat{\mathcal{P}}_{G}$ is the Gutzwiller projector that enforces the mutual hard-core constraint, and $\hat{S}_{j}^{z}$, $\hat{S}_{j}^{\pm} = \hat{S}_{j}^{x} \pm i \hat{S}_{j}^{y}$ are spin-1/2 operators at site $j$. The hole tunnels between neighboring sites with amplitude $t_{\sigma}>0$, and exhibits $1/r_{ij}^3$ decay with euclidean distance; $J_{\perp}$ and $J_{z}$ are the XY and Ising spin coupling strength, respectively, both obeying $1/r_{ij}^6$ decay. We exploit the difference of algebraic decay exponents to access the regime $t\gg J$ by a large lattice spacing~$a$. In the following experiments, atoms are trapped in optical tweezer arrays arranged as either a triangular ladder (Fig.\,\ref{fig:setup}a) or a 2D triangular lattice (Fig.\,\ref{fig:setup}b). For lattice spacing $a=14.7\,\si{\micro\meter}$ and a $46\,\si{G}$ magnetic field applied perpendicular to the array plane, the nearest-neighbor interactions are $t_{\uparrow}\approx t_{\downarrow}\approx 2\pi\times1\,\si{\mega\hertz}$, $J_{\perp}\approx2\pi\times0.08\,\si{\mega\hertz}$ and $J_{z}\approx2\pi\times(-0.05)\,\si{\mega\hertz}$ (see Method, Extended Data Tab.\,\ref{tab:params}).

The experiment follows a three-step sequence: (i) state preparation, (ii) evolution, and (iii) final state measurement (Fig.\,\ref{fig:setup}a). The starting point is the creation of a spin-polarized product state with few localized charge and spin excitations $\ket{\downarrow\cdots\downarrow h \uparrow \downarrow \cdots \downarrow}$~\cite{LS_Bornet_PRL} (see Methods): after initializing each atom in the $\ket{h}$ state, we apply site-dependent addressing light shifts, implementing $\hat{H}_{\uparrow}=\hbar\sum_{i} \delta_{\downarrow}\hat{n}_i^{\downarrow}$ on the sites $i$ to be prepared in the $\ket{\uparrow}$ state, and $\hat{H}_{h}=\hbar\sum_{j} \delta_{\downarrow}\hat{n}_j^{\downarrow}  + \delta_{\uparrow}\hat{n}_j^{\uparrow}$ on the sites $j$ to be prepared in the $\ket{h}$ state. We then sequentially apply microwave pulses resonant with the $\ket{h}\to\ket{\downarrow}$ and subsequently the $\ket{h}\to\ket{\uparrow}$ transitions. The microwaves flip the non-addressed sites to $\ket{\downarrow}$, drive the sites addressed by $\hat{H}_{\uparrow}$ to $\ket{\uparrow}$, while the remaining sites are left in the $\ket{h}$ state. This method enables arbitrary product-state preparation. To create the spin-polarized background, we only address a few sites, leaving the majority of non-addressed sites in the $\ket{\downarrow}$ state. Positive light shifts bring the initial state near the ground state of $\hat{H}_{tJ}+\hat{H}_{\uparrow}+\hat{H}_{h}$. We dynamically prepare a low-energy state of $\hat{H}_{tJ}$ by slowly decreasing the light shifts with exponential ramp profiles $\delta_{\uparrow}(T),\delta_{\downarrow}(T)$~\cite{Braun2013,CSB_Chen_2023,TLL_Emperauger_2025}. Finally, we measure the state by mapping one or two of the three Rydberg levels to the presence of an atom for detection (see Methods).

\section*{One-hole-one-magnon bound states}
\label{sec:1H1M}

We first investigate the kinetically-induced bound state formed by a single hole and a single magnon (1H1M) on a triangular ladder. In a fully spin-polarized single bond, a hole lowers its kinetic energy by forming an antisymmetric wavefunction between the two sites, thereby enhancing its mobility~\cite{OL_Prichard2024} (see also Methods). On a triangular plaquette, however, this antisymmetric condition cannot be satisfied simultaneously across all three bonds, resulting in kinetic frustration. The presence of a magnon resolves this frustration. When the magnon forms an antisymmetric singlet state ($\ket{\uparrow \downarrow} - \ket{\downarrow \uparrow}$) with a neighboring spin-down atom, it effectively flips the sign of the hole's hopping amplitude ($t^{\text{eff}} = -t < 0$), acting as an effective $\pi$-flux. With this negative effective hopping, the hole's kinetic energy is minimized by a symmetric wavefunction, which can be accommodated in the triangular plaquette without frustration. The magnon thus relieves the kinetic frustration, lowering the hole's energy, and consequently they form a stable bound state. The resulting bound state is therefore highly mobile and can propagate through the lattice unhindered by kinetic frustration, as shown in Fig.~\ref{fig:setup}c. We can quantify this binding energy by $E_b = E(1H1M)-E(1H)-E(1M)+E(0H0M)$ such that $E_b<0$ indicates binding, and $E(nHmM)$ is the ground state energy of $n$ holes and $m$ magnons. On triangular ladders, the binding energy depends on the geometry, as shown in Fig.\,\ref{fig:setup}d. As we tune the ratio~$h/a$, we find a large negative binding energy around $E_b=-1.5|t|$ at $h/a=0.5$, which decreases to $E_b=-0.85|t|$ at $h/a=0$ where the ladder becomes a 1D chain with dipolar tunneling.

\begin{figure*}
\mbox{}
\includegraphics[width=\textwidth]{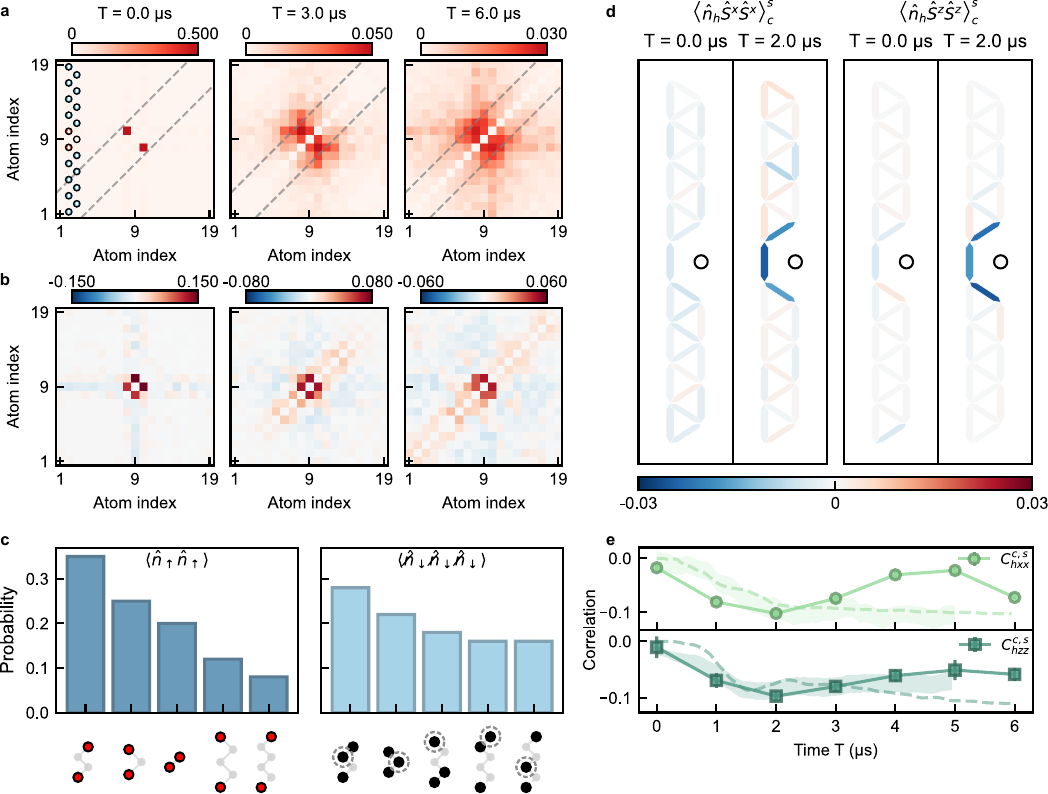}
\caption{\label{fig:1H2M_bound_state} 
{\bf Observation of a 1H2M bound state and kinetically-induced antiferromagnetism.}
\textbf{a,} Time evolution of the magnon-magnon correlation, $\braket{\hat{n}^\uparrow_i \hat{n}^\uparrow_j}$. The system is initialized at $T=0.0\,\mu\text{s}$ with two localized magnons as shown in the inset. As the system evolves, the correlations delocalize across the ladder but remain concentrated near the main diagonal indicating that the two magnons move together as a bound pair. The gray dashed lines indicate the position of strongest correlations of ground state.
\textbf{b,} Hole-magnon connected correlations $\braket{\hat{n}_i^h \hat{n}_j^\uparrow}^{s}_{c}$ at different times.
\textbf{c,} Probabilities of particle configurations at $T=6\,\mu\text{s}$. The left panel displays the probability of finding the two magnons in various relative arrangements, as measured in the $\uparrow$ basis, confirming their tendency to remain close. The right panel plots the three-body correlator $\braket{\not{\hat{n}}_{\downarrow}\not{\hat{n}}_{\downarrow}\not{\hat{n}}_{\downarrow}}$, measuring the likelihood of different configurations of the 1H2M state. The hole's position (dashed circle) is inferred by comparing these configurations with the two-magnon arrangements shown in the left panel.
\textbf{d,} Spatial map of hole-induced magnetic order. The plots show the connected three-body hole-spin-spin correlators for the transverse, $\braket{\hat{n}^h \hat{S}^x \hat{S}^x}_c^s$ (left), and longitudinal, $\braket{\hat{n}^h \hat{S}^z \hat{S}^z}_c^s$ (right), components. At $T=2.0\,\mu\text{s}$, strong antiferromagnetic correlations (blue links) develop on the bonds around the hole's initial location (red dot), demonstrating the kinetically-induced binding requires formation of singlets around the hole.
\textbf{e,} Time evolution of the nearest-neighbor hole-spin-spin correlators in the hole frame, $C_{hxx}^{c,s}(\mathbf{d}_i,\mathbf{d_j})$ (transverse) and $C_{hzz}^{c,s}(\mathbf{d}_i,\mathbf{d_j})$ (longitudinal), summed over the sites adjacent to the hole. Dashed (shaded) lines indicate ideal simulation (simulation including experimental imperfections). Error bars denote one standard deviation extracted via bootstrap.
}
\end{figure*}

To prepare the 1H1M bound state, we rely on the conservation of the number of holes and magnons during the evolution. We initialize the product state $|\downarrow\cdots\downarrow h \uparrow \downarrow \cdots \downarrow\rangle$, as shown in Fig.\,\ref{fig:1H1M_bound_state}a. Then, we connect it to low-energy states by reducing the light shifts (see Methods). A hallmark of binding is provided by the connected correlations between its constituents: $\braket{\hat{n}_i^h \hat{n}_j^\uparrow}_c = \braket{\hat{n}_i^h \hat{n}_j^\uparrow} - \braket{\hat{n}_i^h}\braket{\hat{n}_j^\uparrow}$. Although our current measurement method is not able to distinguish all three different states in one single shot, we can still statistically reconstruct observables with permutation symmetry~\cite{Greiner2016} (see also Methods):
\begin{equation}
\label{eq:sym_connected_2body_corr}
    \braket{\hat{n}_i^h \hat{n}_j^\uparrow}_{c}^{s} =\left( \braket{\hat{n}_i^h \hat{n}_j^\uparrow}_c + \braket{ \hat{n}_i^\uparrow \hat{n}_j^h}_c \right) / 2.
\end{equation}
In Fig.\,\ref{fig:1H1M_bound_state}b, we plot the correlation map at different times~$T$ during and after the ramp. The two positive correlations of the map at $T=0\,\mu\text{s}$ reflect the initial state with the hole and the magnon at sites 8 and 9. As $\delta_\uparrow$ and $\delta_\downarrow$ are decreased, the hole-magnon pair delocalizes along the ladder and correlations start to spread. The positive correlations on the super- and sub-diagonals indicate that the hole and magnon mainly reside on neighboring sites, but acquire mobility as a composite object, a direct consequence of binding. The final state at $T=4\,\mu\text{s}$ does show a bound pair. The higher correlations around the site 9 indicates that this bound pair tends to be more localized at the center of the ladder. We quantify this by computing the center of mass (COM) distribution, defined as
\begin{equation}
    \label{eq:com_def}
    \mathcal{P}_{\text{COM}}(\mathbf{d})=\sum_{\substack{(\mathbf{r}_i+\mathbf{r}_j})/2=\mathbf{d}}\braket{\hat{n}_i^h \hat{n}_j^\uparrow}^s 
\end{equation}
between any pair of sites with position $\mathbf{r}_i$ and $\mathbf{r}_j$, whose COM locates at $\mathbf{d}$. Fig.\,\ref{fig:1H1M_bound_state}c shows a cut of this distribution along the midline between the two legs (see inset), corresponding to having the hole and the magnon on different legs. We compare it to the theoretical wavefunction, of a particle in a box, $\sin^2y$, which shows that the distribution is well-captured by modeling the bound pair as single particle confined within the ladder.

Next, we characterize the existence of the kinetically-induced bound state in a 2D triangular lattice. We probe the formation of the 1H1M composite by initializing the pair on adjacent central sites of a triangular lattice, as shown in Fig.\,\ref{fig:1H1M_bound_state}d, and applying a similar ramp as for the ladder experiments. After an evolution of $T=5\,\mu\text{s}$, we measure the symmetrized and connected up-hole correlation versus relative separation between the two particles, as shown in Fig.\,\ref{fig:1H1M_bound_state}e,
\begin{equation}
    \label{eq:connected_correlations_2D_holeframe}
    C^{c,s}(\mathbf{d})=\frac{1}{\mathcal{N}_{\mathbf{d}}}\sum_{\mathbf{r}_{i}-\mathbf{r}_{j}=\mathbf{d}}\langle \hat{n}_i^h \hat{n}_j^\uparrow \rangle_{c}^{s},
\end{equation}
where $\mathcal{N}_{\mathbf{d}}$ is the number of possible pairs connected by a vector $\mathbf{d}$. Figure\,\ref{fig:1H1M_bound_state}e reveals a strong positive correlation confined to nearest-neighbors, surrounded by negative correlations, showing that the hole and magnon stay bound. We further compute the COM spatial distribution (Eq.\,\eqref{eq:com_def}) as shown in Fig.\,\ref{fig:1H1M_bound_state}f. The measured probability delocalizes away from its initial position, indicating that the composite object moves freely. This provides  evidence of a mobile, kinetically-induced bound pair in the 2D triangular lattice.

To analyze the binding length, we average the symmetric non-connected up-hole correlations, computed in a similar way to the connected ones in Eq.\,\eqref{eq:sym_connected_2body_corr}, over the pairs of sites separated by the same distance~$d$:

\begin{equation}
    \label{eq:sym_ns_2body_corr_avg_d}
    C^{s}(d)=\frac{1}{\mathcal{N}_d}\sum_{d=|i-j|}\braket{ \hat{n}_i^h \hat{n}_j^\uparrow}^{s}
\end{equation}
where $\mathcal{N}_d$ is the number of possible pairs of sites separated by a distance $d$. The distance metric,~$|\!\cdot\!|$, depends on the geometry: for the ladder, distance is defined as the absolute difference of the sites' index (see Fig.\,\ref{fig:1H1M_bound_state}a); for the 2D lattice, we use the Manhattan distance in units of lattice spacing. 

Fig.\,\ref{fig:1H1M_bound_state}g shows how correlations evolve with distance in the ladder (top) and 2D (bottom). The exponential decay originates from the tight binding of the two particles and can be understood as the evanescent part of the bound pair wave function analogous to the wavefunction outside a finite potential well. The extracted decay rates are in both dimensions slower than the theoretical ones of the ground state. We explain this discrepancy by two factors: a diabatic component of our ramp preparing bound states with higher energy, and a noise floor originating from state-preparation-and-measurement errors (see Methods). This is also confirmed by ab-initio numerical simulations including experimental imperfections for the ladder case, which well matches the experimental data. Although the binding energies are similar in ladder and 2D array, the bound state is theoretically tighter in 2D. This dimensional effect can be understood by considering how the magnon-hole pair relieves kinetic frustration. While a nearest-neighbor pair relieves frustration on two triangular plaquettes in both dimensions, separating the pair by an additional site in 2D offers no energetic benefit, unlike in the ladder. Consequently, the energy penalty for separation is greater in 2D, leading to a more compact bound state, as shown in the inset of Fig.\,\ref{fig:1H1M_bound_state}g. However, the experimental imperfections mentioned above prevent us from obtaining a clear difference in the measured decay rates in the non-connected correlations. Nevertheless, this dimensional effect can be observed in the connected correlation maps and the bottom panel of Fig.\,\ref{fig:1H1M_bound_state}g, as computation of connected correlations removes a part of the uncorrelated errors and thus improves the signal. Indeed, connected correlations between next-nearest neighbors are already negative in the 2D case but remain positive at the center of the ladder. 

\begin{figure*}
\mbox{}
\includegraphics[width=\textwidth]{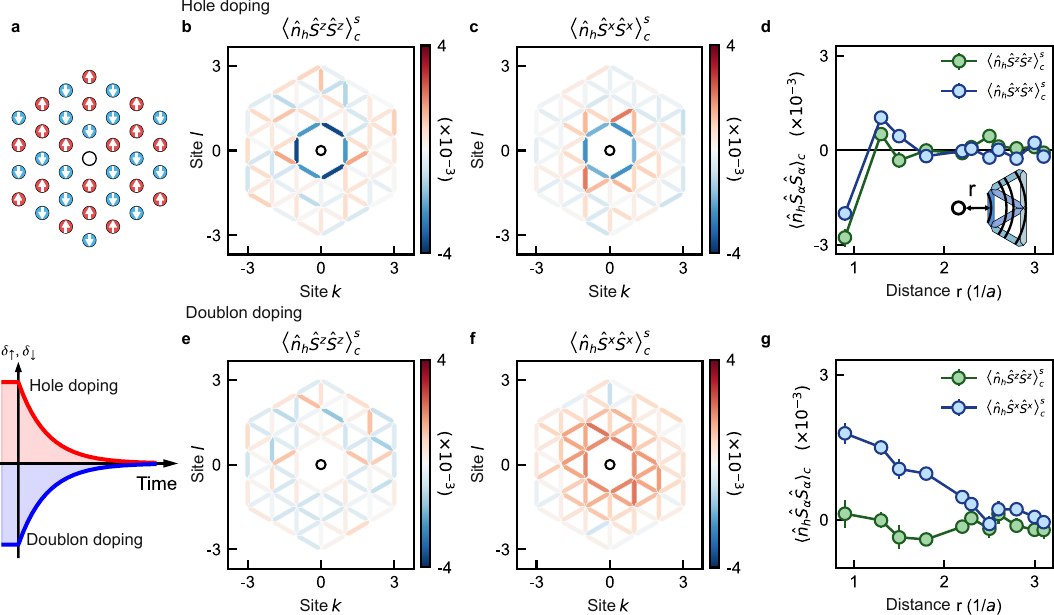}
\caption{\label{fig:2d_bound_state} 
{\bf Spin bag induced by mobile dopants in a 2D triangular lattice.}
\textbf{a,} The top panel shows the initial state of the 2D triangular lattice, which can be doped with a single hole or particle (white circle) in a background of spin-up (red) and spin-down (blue) atoms. The bottom panel illustrates the quasi-adiabatic protocol, showing the time-dependent light shift used to create a hole (positive values, red) or a particle (negative values, blue). The correlations are measured at $T=2\,\mu\text{s}$.
\textbf{b, c,} Spatial maps of connected three-body correlators for hole doping in the hole frame, where $k,l$ are the index of sites. Both the longitudinal correlator $\braket{\nhole \Sz \Sz}_c^s$ (\textbf{b}) and the transverse correlator $\braket{\nhole \Sx \Sx}_c^s$ (\textbf{c}) reveal strong antiferromagnetic correlations (blue) on the bonds surrounding the hole.
\textbf{d,} Radially averaged correlators for hole doping, plotted as a function of distance $r$ from the hole. Here $r$ is defined as the distance between the hole and the center of bonds, as depicted in the inset.
\textbf{e, f,} Spatial maps of correlators for a doublon dopant in the particle frame. The particle induces strong ferromagnetic transverse correlations $\braket{\nhole \Sx \Sx}_c^s$ (\textbf{f}), while the longitudinal correlations $\braket{\nhole \Sz \Sz}_c^s$ (\textbf{e}) are significantly suppressed.
\textbf{g,} Radially averaged correlators for doublon doping, highlighting the transverse ferromagnetic correlations and a weak longitudinal correlations. Error bars denote one standard error estimated via bootstrap, and are smaller than marker size.}
\end{figure*}

\section*{Multi-particle bound states}
\label{sec:multi_particle_bound_state}

Increasing the density of spin excitations can lead to the formation of multi-particle bound states~\cite{Attraction2024}. With two magnons and one hole, the low-energy spectrum contains three-particle bound states. The largest negative binding energy of the one-hole-two-magnon (1H2M) bound state is obtained for a triangular ladder with a ratio of $h/a=0.5$, as indicated in Fig.\,\ref{fig:setup}c, therefore we decrease the ratio $h/a$ to reach a similar binding energy as in the 1H1M case.

To prepare this state, we initialize a spin-polarized configuration, $\ket{\downarrow\cdots\downarrow \uparrow h \uparrow \downarrow \cdots \downarrow}$ (see Fig.\,\ref{fig:1H2M_bound_state}a inset), and then ramp down the light shifts. The existence of the 1H2M bound state is first probed by measuring magnon-magnon correlations, $\braket{\hat{n}_i^\uparrow \hat{n}_j^\uparrow}$, as shown in Fig.\,\ref{fig:1H2M_bound_state}a. After state preparation, the two magnons are next-nearest neighbors, appearing as two spots in the correlation map. As the system approaches its low-energy state, the magnons delocalize in tandem along the ladder while remaining in close proximity, indicated by the spreading parallel to the main diagonal of the correlation map.

Next, we measure the connected symmetrized hole-magnon correlation, $\braket{\hat{n}_i^h \hat{n}_j^\uparrow}^{s}_{c}$, at various times~$T$ (Fig.\,\ref{fig:1H2M_bound_state}b). The hole remains tightly bound with the magnons, but in contrast to the 1H1M case, the 1H2M correlation map occupies four diagonal lines instead of two. This is compatible with the presence of a second magnon moving together with the hole-magnon bound state. 

To investigate the internal structure of the 1H2M bound state, we analyze the full counting statistics at time $T=6\,\mu\text{s}$ to determine the probability of specific particle configurations. For measurements in the $\ket{\uparrow}$-basis, we record the positions of the magnons and compute their relative arrangement. The histogram of the magnon-magnon configurations are shown in the left panel of Fig.\,\ref{fig:1H2M_bound_state}c. When two magnons are adjacent, they cannot relieve kinetic frustration by forming a singlet, therefore the most probable arrangement has the two magnons separated by three sites. The probability of a given configuration decays quickly as the distance between the two magnons increases. We then measure the operator $\hat{\not{n}}_{\downarrow}\hat{\not{n}}_{\downarrow}\hat{\not{n}}_{\downarrow}$ in the $\ket{\downarrow}$-basis. The right panel of Fig.\,\ref{fig:1H2M_bound_state}c shows a histogram of these results, where black circles represent the missing atoms. Although cannot distinguish between a hole and a magnon, we can infer the hole's position by comparing these configurations with $\ket{\uparrow}$-basis measurements and assigning the "extra defect" as the hole (labeled with a dashed circle). The dominant three-body configuration features the hole adjacent to one magnon and the other magnon is separated by a single site from the 1H1M pair.

Next, we investigate the formation of singlet correlations near the hole. As illustrated in Fig.\,\ref{fig:setup}(c), the binding between magnons and the hole is driven by increased hole mobility. This mechanism couples the magnons to the polarized spin background, forming singlets around the hole. To probe this local magnetic order, we first measure the connected symmetrized three-body hole-spin-spin correlator $\braket{\hat{n}_{\mathbf{r}_{0}}^{h}\hat{S}^{\alpha}_{\mathbf{r}_{0}+\mathbf{d}_{i}}\hat{S}^{\alpha}_{\mathbf{r}_{0}+\mathbf{d}_{j}}}_{c}^{s}$, which evaluates spin correlations near the hole's position, where $\alpha\in\{x,z\}$ is the spin component, $\mathbf{r}_0$ is the initial position of the hole, and $\braket{\cdots}_c^s$ denotes the connected, symmetrized correlator (see Methods). Initially, the system is in a product state with no correlations, as verified for both the $x$- and $z$-directions in Fig.\,\ref{fig:1H2M_bound_state}d. As the light shifts are ramped down, the hole delocalizes and induces strong antiferromagnetic correlations in its vicinity. At $T=2\,\mu\text{s}$, the bonds adjacent to the hole exhibit pronounced antiferromagnetic correlations in both the $x$- and $z$-direction with nearly equal strength, indicating the local singlet formation. The confinement of these correlations to the hole's nearest-neighborhood confirms that the magnons are bound to it. For two magnons, this antiferromagnetic pattern distributes over the four nearest sites, as each magnon is entangled with a nearby down spin to form a singlet. We quantify the emergence of correlations by tracking the time evolution of the hole-spin-spin correlator in the hole frame $C^{c,s}_{h\sigma\sigma}(\mathbf{d}_i,\mathbf{d}_j)=\sum_{\mathbf{r}}\braket{\hat{n}_{h,\mathbf{r}}\hat{S}^{\sigma}_{\mathbf{r}+\mathbf{d}_{i}}\hat{S}^{\sigma}_{\mathbf{r}+\mathbf{d}_{j}}}_{c}^{s}$, where the summation is over all hole positions. Fig.\,\ref{fig:1H2M_bound_state}e shows the time evolution of the nearest-neighbor sites around hole (see inset). Starting from nearly zero at time $T=0$, both the transverse ($\braket{\hat{n}^h \hat{S}^x \hat{S}^x}_c^s$) and longitudinal ($\braket{\hat{n}^h \hat{S}^z \hat{S}^z}_c^s$) correlations rapidly develop, reaching their largest negative value at $T=2\,\mu\text{s}$. This demonstrates a quick formation of singlets as the hole-magnon bound state forms. Subsequently, the magnitude of the correlations slowly decreases, which may be attributed to the finite lifetime of the prepared many-body state. Ideal simulations including the ramp profile show no late-time decay; thus, the observed decay likely arises from finite initial entropy of the initial state and experimental imperfections, which is captured by our \textit{ab-initio} numerical simulations including errors (see Methods).

\section*{Spin bag in 2D arrays}
\label{sec:2D}
Finally, we investigate kinetic magnetism in the spin-balanced 2D triangular lattice in our bosonic \tJ~quantum simulator. The motion of a dopant through the lattice is expected to dress it with a cloud of spin correlations, forming a composite object known as a magnetic polaron or spin bag~\cite{Attraction2024}. To probe the nature of the dressing cloud, we initialize the system with a single dopant at the center of a spin-balanced background containing 18 spin-up and 18 spin-down atoms (Fig.\,\ref{fig:2d_bound_state}a), and subsequently ramp the light shifts to zero. We can tune the sign of the light shift to target either the ground state or the highest excited state of the \tJ Hamiltonian, which effectively reverses the sign of tunneling~$t$. A positive light shift (red curve in Fig.\,\ref{fig:2d_bound_state}a) prepares the kinetically-frustrated ground state with $t>0$, analogous to the \textit{single} hole doping case of fermionic \tJ model (see Methods). In contrast, a negative light shift (blue curve) prepares a negative temperature state~\cite{Braun2013}, which is equivalent to the ground state of an unfrustrated Hamiltonian with $t<0$ and is equivalent to a \textit{single} doublon dopant in the fermionic \tJ model. This allows us to probe how these distinct dopant types generate different magnetic orders~\cite{CSB_Chen_2023}.

\begin{figure}
\mbox{}
\includegraphics[width=0.48\textwidth]{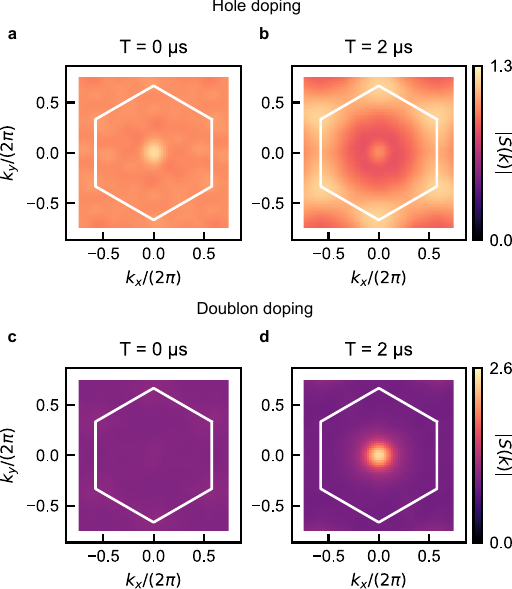}
\caption{\label{fig:2d_structure_factor} 
{\bf Kinetic magnetism.}
Experimentally measured structure factor $|S(\mathbf{k})|$ of two-body correlation $\braket{\hat{S}^x_i\hat{S}^x_j}_c$ for hole doping (\textbf{a, b}) and doublon doping (\textbf{c, d}), where $\mathbf{k}$ is the momentum. The white hexagon outlines the first Brillouin zone. \textbf{a,} For hole doping at $T=0\,\mu\text{s}$, the initial state shows weak magnetic correlations. \textbf{b,} After evolving for $T=2\,\mu\text{s}$, correlations emerge at the corners of the Brillouin zone (K points), signaling the formation of $120\degree$ antiferromagnetic order induced by the mobile hole. \textbf{c,} For doublon doping at $T=0\,\mu\text{s}$, the system has negligible magnetic correlations. \textbf{d,} By $T=2\,\mu\text{s}$, a strong peak develops at the center of the Brillouin zone ($\Gamma$ point), indicating the emergence of ferromagnetic order driven by the particle's motion.
}
\end{figure}

We characterize the induced magnetic order by measuring the connected three-body dopant-spin-spin correlators in the dopant's frame, $C^{c,s}_{h\alpha\alpha}(\mathbf{d}_i,\mathbf{d}_j)$, for both longitudinal ($\alpha=z$) and transverse ($\alpha=x$) spin components. For hole doping, shown in Fig.\,\ref{fig:2d_bound_state}(b, c), the spatial maps reveal strong negative (antiferromagnetic) correlations, indicated by blue links, on the bonds nearest to the hole. This AFM ordering is present for both the longitudinal $C^{c,s}_{hxx}$ and transverse $C^{c,s}_{hzz}$ correlators. This result reveals the hole's dressing by cloud of antiferromagnetic order. Furthermore, we observe weaker positive (ferromagnetic) correlations on the next-nearest-neighbor bonds, a  signature of $120\degree$ antiferromagnetic order in a triangular lattice. This demonstrates that to enhance its mobility, the mobile hole locally structures the surrounding spins into the $120\degree$ ordered phase, a phenomenon consistent with Haerter–Shastry antiferromagnetism~\cite{HS2005}. The radially averaged data in Fig.\,\ref{fig:2d_bound_state}d quantitatively confirms this, showing that the nearest-neighbor correlations are strongly antiferromagnetic and have a similar magnitude for both spin components.

In sharp contrast, a doublon dopant induces ferromagnetic magnetic order. In this regime, the tunneling is unfrustrated ($t<0$), leading to Nagaoka ferromagnetism. The spatial maps in Fig.\,\ref{fig:2d_bound_state}f show that a doublon dopant generates strong ferromagnetic correlations, indicated by red links. This ferromagnetic dressing is strikingly anisotropic in the spin direction: it is strong in the transverse, $x$ direction, but is almost entirely absent in the longitudinal, $z$, direction. The radially averaged correlators in Fig.\,\ref{fig:2d_bound_state}g highlight this disparity: strong finite-range positive transverse correlations, while the longitudinal response remains close to zero. Hence, the observed spin bag around the particle is polarized within the XY plane. 
We attribute this anisotropy to our state preparation protocol, in which the initial state is antiferromagnetic and polarized along the $z$-direction, breaking the $SU(2)$~symmetry. Starting from an eigenstate of $\hat{S}_{\text{tot}}^z = \sum_j \hat{S}^z_j$ -- which is conserved throughout the dynamics -- developing positive longitudinal correlations around the hole would require a substantial rearrangement of the spins. However, the spins initially exhibit strong $x$~fluctuations, facilitating the formation of ferromagnetic correlations in the XY plane. Quantifying the role of the initial state in the anisotropic correlations is an interesting subject and requires further study.

We further investigate the bosonic dopant induced spin magnetism by computing the spin structure factor in the transverse direction, $S(\mathbf{k}) = \frac{1}{N} \sum_{i,j} e^{i\mathbf{k}\cdot(\mathbf{r}_i - \mathbf{r}_j)} \braket{\hat{S}^x_{\mathbf{r}_i} \hat{S}^x_{\mathbf{r}_j}}_c$, for both dopant types, as shown in Fig.\,\ref{fig:2d_structure_factor}. In the case of hole doping, the system initially shows weak magnetic correlations at $T=0\,\mu\text{s}$ (Fig.\,\ref{fig:2d_structure_factor}a). As the system evolves to $T=2\,\mu\text{s}$, peaks develop at the corners of the Brillouin zone (the K points), as seen in Fig.\,\ref{fig:2d_structure_factor}b. The emergence of these peaks signals $120\degree$ antiferromagnetic order. In contrast, the doublon dopant induces a completely different magnetic structure. While the initial state is also a paramagnet (Fig.\,\ref{fig:2d_structure_factor}c), an evolution of $T=2\,\mu\text{s}$ leads to the formation of a single, dominant peak at the $\Gamma$ point, as shown in Fig.\,\ref{fig:2d_structure_factor}d, which is signals ferromagnetic order consistent with the transverse ferromagnetic spin bag around the doublon measured in Fig.\,\ref{fig:2d_bound_state}f. These measurements thus reveal how different dopant types dictate the emergent magnetic order, providing a direct view of kinetic magnetism induced by bosonic dopants in the triangular lattice.

\section*{Conclusion}

In conclusion, we have reported the first direct experimental imaging of kinetically-induced bound states using a Rydberg-based quantum simulator of the frustrated bosonic \tJ model. By precisely preparing and tracking individual holes and magnons on both ladders and 2D arrays, we have demonstrated the formation of mobile one-hole-one-magnon and more complex three-particle one-hole-two-magnon bound states. The persistence of strong spatial correlations between the constituent particles, combined with the delocalization of their center of mass, provides clear evidence for these composite quasiparticles.

Crucially, we have provided microscopic insight into the binding mechanism itself. Our measurements of three-body hole-spin-spin correlators reveal that the mobile hole dynamically dresses itself with a "spin bag" of local antiferromagnetic correlations. This induced magnetic environment relieves the inherent kinetic frustration on the frustrated triangular lattice, lowering the system's kinetic energy and creating an effective attraction between the hole and the magnons.

Furthermore, we have shown that the nature of this magnetic dressing is dopant-dependent. In the 2D triangular lattice, a mobile hole induces $120\degree$ antiferromagnetic order, whereas a mobile particle induces distinct transverse ferromagnetic correlations. This work opens a new avenue for exploring many-body phenomena driven by kinetic frustration, including the formation of quantum many-body states characterized by higher-order correlation functions, as discussed in the Fermi-Hubbard model on the triangular lattice \cite{he2023itinerantmagnetismtriangularlattice,morera2024itinerantmagnetismmagneticpolarons,Chen2025}. Additionally, the relation to models with beyond nearest-neighbor tunnelings could be explored~\cite{ttprimeJ2025,Zhang2025_KF}. Future studies could investigate the interactions and collisions between these composite particles, a crucial step toward understanding mechanisms of unconventional pairing driven by kinetic frustration. Spectroscopic probes analogous to those proposed in \cite{Exploring2024} could also be employed to directly measure the binding energy of these composite bound states. The programmability of our quantum simulator provides a powerful platform for systematically studying these exotic states of matter, bridging the gap between theoretical models of strongly correlated systems and condensed matter experiments.

\bibliographystyle{unsrt}  
\bibliography{main}

\newpage

\subsection*{Methods}

\noindent{\bf Mapping the \tJ model.}
The experiments were performed with arrays of $^{87}$Rb atoms trapped in optical tweezers, using the setup described in previous works~\cite{tJ_qiao2025}. We map the \tJ model onto three Rydberg states: $\ket{\downarrow} \coloneqq \ket{S} = \ket{60S_{1/2}, m_{J} = 1/2}$, $\ket{h} \coloneqq \ket{P} = \ket{60P_{3/2}, m_{J} = 1/2}$, $\ket{\uparrow} \coloneqq \ket{S'} = \ket{61S_{1/2}, m_{J} = 1/2}$. We use a $46\,\si{G}$ magnetic field orthogonal to the atomic plane to ensure isotropic interactions and to isolate the effective spin states from other Zeeman levels. The states involved during the Rydberg sequence and their resonance frequencies are given in Extended Data Fig.\,\ref{fig:spectrum}.

\begin{extfig}
\mbox{}
\includegraphics[width=0.3\textwidth]{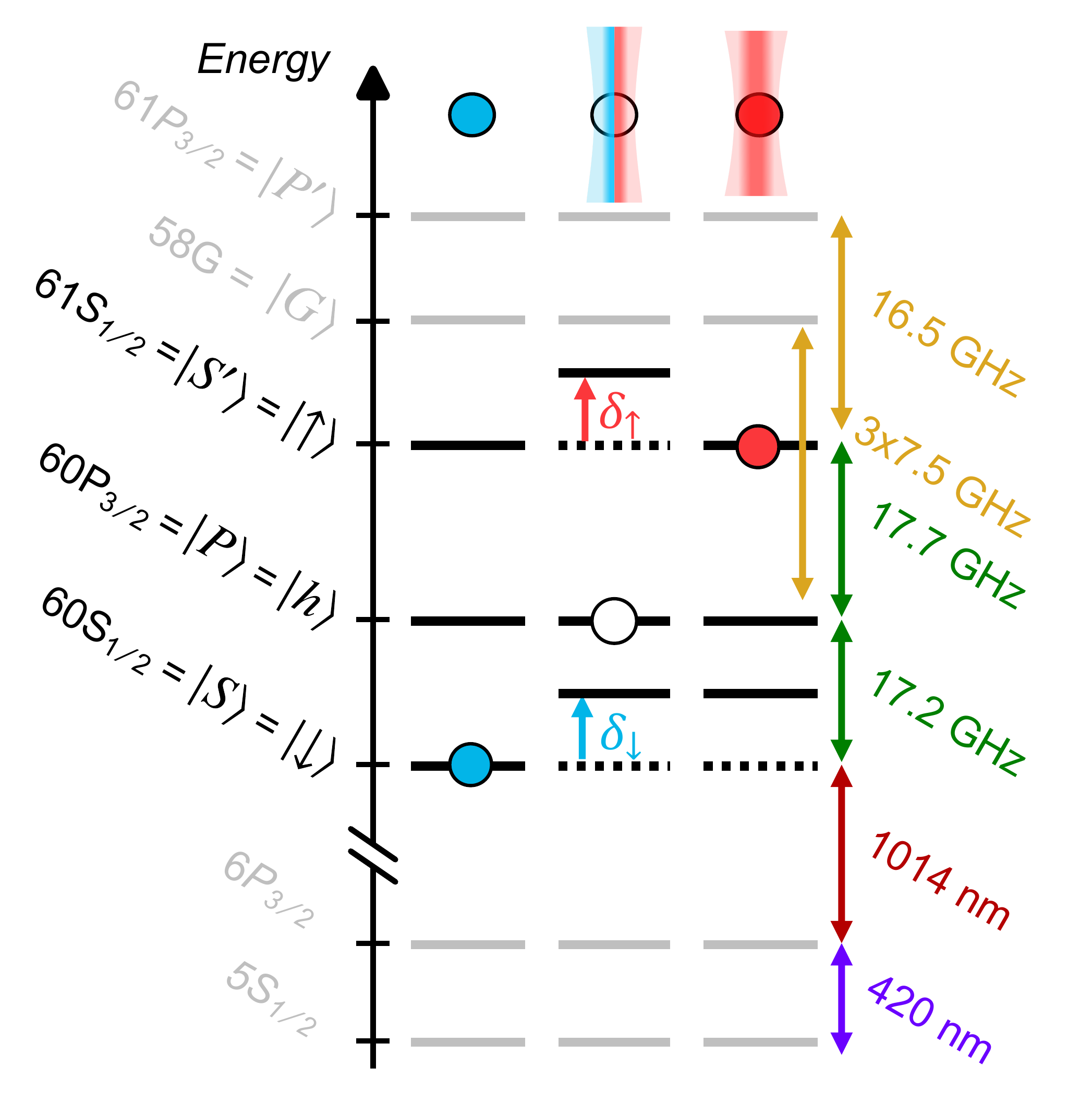}
\caption{\label{fig:spectrum} 
\textbf{Involved states and transitions} The states used in the mapping of the \tJ Hamiltonian are indicated in black. Each of the three columns represents one class of atoms: non-addressed atoms (left column)
are prepared in $\ket{\downarrow}$, atoms with both $\delta_\downarrow$ and $\delta_\uparrow$ light shifts (center column) are prepared in $\ket{h}$, and atoms with only the $\delta_\downarrow$ light shift (right
column) are prepared in $\ket{\uparrow}$.}

\end{extfig}

The resonant dipole exchange and van-der-Waals interactions between atoms in the above  Rydberg states give rise to the bosonic $t$--$J$ Hamiltonian:
\begin{equation}
\label{_Hamiltonian}
\begin{aligned}
&\hat{H}_{tJ} = \hat{H}_{t}+\hat{H}_{J}\ ,  \\
&\hat{H}_{t} =+\hbar\sum_{i<j}\sum_{\sigma=\downarrow,\uparrow
}t_{\sigma}\frac{a^3}{{r}_{ij}^3}\hat{\mathcal{P}}_{G}\left(\hat{b}_{i,\sigma}^\dagger\hat{b}_{j,\sigma}+\text { h.c. }\right)\hat{\mathcal{P}}_{G}, \\
&\hat{H}_{J} = \hbar\sum_{i<j}\frac{a^6}{{r}_{ij}^6}\left[J_{z}\hat{S}_i^z \hat{S}_j^z+\frac{J_{\perp}}{2}\left(\hat{S}_i^{+} \hat{S}_j^{-}+\text {h.c. }\right)\right]\ , 
\end{aligned}
\end{equation}
with hopping terms $t_{\downarrow}, t_{\uparrow}$ generated by resonant dipole interactions with the typical scaling $1/r^3$ with distance. The second order dipole exchange gives rise to $J_{\perp}$. The van-de-Waals interactions correspond to $J_{z}$. The two latest terms feature a $1/r^6$ scaling.

\begin{table*}[ht]
    \label{table:interactions}
    \centering
    \begin{tabular}{cccccccccc}
    \toprule
    \textbf{$h/a$} & 
    \textbf{$a$ (\si{\micro\meter})} &
    \multicolumn{8}{c}{\textbf{Interaction strengths} ($2\pi\times$MHz)} \\
    &
    &
    \multicolumn{4}{c}{\textbf{Rung}} &
    \multicolumn{4}{c}{\textbf{Leg}} \\
    \cmidrule(lr){3-6}\cmidrule(lr){7-10}
    &
    & $t_{\uparrow}$ & $t_{\downarrow}$ & $J_{\perp}$ & $J_{z}$ & $t_{\uparrow}$ & $t_{\downarrow}$ & $J_{\perp}$ & $J_{z}$ \\
    \midrule
    $\sqrt{3}/2$ & 14.7 & $1.1$ & $1.2$ & 0.107 & $-0.071$ & $0.95$ & $1.0$ & 0.075 & $-0.05$ \\
    0.5 & 19.6 & $1.1$ & $1.2$ &  0.110 & $-0.076$ & $0.40$ & $0.42$ & 0.013 & $-0.008$ \\
    \bottomrule
    \end{tabular}
    \caption{\label{tab:params}
    Rung and leg interactions for different rung-to-leg ratios $h/a$. All interaction strengths are given in units of $2\pi \times \si{\MHz}$. Both the rung and leg use the same notation for the hopping $t$, transverse exchange $J_{\perp}$, and longitudinal exchange $J_{z}$.
    }
\end{table*}

We exploit the different scaling of the interactions with distance ($t \propto 1/r^3$ vs. $J \propto 1/r^6$) to access the kinetically-dominated regime ($t \gg J$) by increasing the lattice spacing $a$. While exploring different coupling regimes by varying the ladder's aspect ratio $h/a$, we adjust $a$ to keep the hopping strength in a range that limits experimental errors during state preparation and readout (see below). Specifically, the one-hole-one-magnon (1H1M) and spin bag experiments were performed on equilateral triangular ladder and lattices with $a=14.7\,\si{\micro m}$. For the one-hole-two-magnon (1H2M) experiments, we use $h/a=0.5$, and increase the spacing to $a=19.6\,\si{\micro m}$ to maintain a rung hopping strength of $t_{\text{rung}} \approx 2\pi\times 1\,\si{MHz}$. This adjustment simultaneously reinforces the $t \gg J$ condition. The resulting interaction parameters are detailed in Extended Data Table.\,\ref{tab:params}.
\\
\\

\noindent{\bf State initialization.}
We load single atoms from a cloud of \Rb atoms in magneto-optical-trap into optical tweezers. The atoms are rearranged into a defect-free array with the desired geometry using a 2D acousto-optic deflector (AOD). After the rearrangement, we sequentially use optical molasses and Raman sideband cooling to lower the temperature of the atoms. Then the atoms are optically pumped to $\ket{g} \coloneq \ket{5S_{1/2}, F=2, m_{F}=2}$ via a $\sigma^{+}$-polarized 795~nm laser. Before the Rydberg excitation, the tweezer depth is adiabatically ramped down by a factor $\sim$~100 to reduce the momentum dispersion of the atomic wavefunctions. We then switch off the tweezers, and use a two-photon stimulated Raman adiabatic passage (STIRAP) with 420~nm and 1014~nm lasers to excite all atoms to $\ket{\downarrow}$.

Extended Data Fig.\,\ref{fig:sate_init} shows the sequence used to prepare the initial spin-polarized state and the subsequent quasi-adiabatic evolution to connect to the low-energy states of the Hamiltonian. The preparation of product states relies on a combination of global microwave pulses (with a Gaussian temporal envelope) and site-dependent light shifts. A $1014\,\si{\nano\meter}$ laser, red-detuned by $\Delta \sim 2\pi \times 300\,\si{\mega \hertz}$ from the $\ket{6P_{3/2}} \longleftrightarrow \ket{60S_{1/2}}$ transition, generates a light shift of $\delta_\downarrow \approx 2\pi \times 25\,\si{\mega\hertz}$. To achieve local control, this laser beam is patterned using a spatial light modulator (SLM), which projects a custom light intensity profile onto the atom array. A second laser, also red-detuned by $2\pi \times 300\,\si{\mega \hertz}$ from the $\ket{6P_{3/2}} \longleftrightarrow \ket{61S_{1/2}}$ transition, creates another light shift, $\delta_\uparrow \approx 2\pi \times 25\,\si{\mega\hertz}$. This second beam is controlled using an Acousto-Optic Modulator (AOM).

\begin{extfig}
\mbox{}
\includegraphics[width=0.45\textwidth]{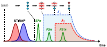}
\caption{\label{fig:sate_init} 
{\bf State initialization and quasi-adiabatic ramp.}
After Rydberg excitation using STIRAP, we apply a sequence of microwave pulses combined with site-resolved light-shifts $\delta_\downarrow$ and $\delta_\uparrow$ to prepare the desired initial product state. 
Both light-shifts are then ramped down to prepare a low energy state.
}
\end{extfig}

Our state preparation protocol relies on a classification of atoms into three groups based on the light shifts they experience: sites intended for magnons ($\ket{\uparrow}$) are addressed by the $\delta_\downarrow$ shift; sites intended for holes ($\ket{h}$) are addressed by both $\delta_\downarrow$ and $\delta_\uparrow$; the remaining unaddressed atoms form the $\ket{\downarrow}$ background. The sequence begins after all atoms are prepared in the $\ket{\downarrow}$ state. First, a global $20\,\si{\nano\second}$ microwave $\pi$-pulse transfers the entire array to the $\ket{h}$ state. Next, the $\delta_\downarrow$ light shift is applied to the target magnon and hole sites. A subsequent $39\,\si{\nano\second}$ $\pi$-pulse, resonant on the $\ket{h} \leftrightarrow \ket{\downarrow}$ transition, returns the unaddressed atoms to their final $\ket{\downarrow}$ state; the addressed atoms remain in $\ket{h}$ as the light shift makes the microwave pulse off-resonant. Finally, the $\delta_\uparrow$ shift is also applied to the designated hole sites, and a $75\,\si{\nano\second}$ $\pi$-pulse resonant on the $\ket{h} \leftrightarrow \ket{\uparrow}$ transition is sent. This final pulse resonantly drives the atoms experiencing only the $\delta_\downarrow$ shift to the $\ket{\uparrow}$ state, while atoms subjected to both shifts are left in the hole state $\ket{h}$.
\\

\begin{extfig}
\mbox{}
\includegraphics[width=\linewidth]{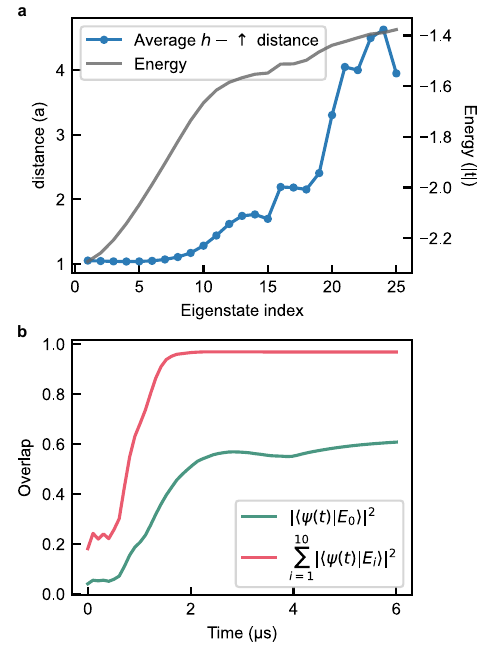}
\caption{\label{fig:overlap}
{\bf Eigenstates overlap during the evolution.} \textbf{a,} The average distance between the hole and the magnon (blue) and the eigenenergy (grey) are shown as a function of the eigenstate index up to $25$. We consider the lowest $10$ eigenstates as bound states, characterized by a short hole-magnon distance. \textbf{b,} The time evolution of the overlap of the prepared state with the true ground state $|E_0\rangle$ (green) and with the subspace spanned by the bound states (red). While the final overlap with the ground state is approximately 60\%, the overlap with the bound-state manifold is close to unity, confirming the successful preparation of the composite particle.}
\end{extfig}

\noindent{\bf{Quasi-adiabatic evolution.}} 
After preparing the target product state, we ramp down both light shifts toward zero to prepare the low-energy states. We first rapidly decrease the light shift to $\sim 2\pi \times 5\,\si{\mega \hertz}$ over $1\,\mu\text{s}$. This initial ramp remains adiabatic because the remaining light shift is still sufficient to suppress dynamics driven by the bare \tJ Hamiltonian. Then both light shifts are exponentially decreased with decay time-constant of $1\,\mu\text{s}$ using an AOM. The finite evolution time of our experimental ramp prevents the perfect preparation of the ground state. In an ideal simulation of our ramp profile, the overlap of the final state with the true ground state is around $60\%$, as shown in Extended Data Fig.\,\ref{fig:overlap}b, indicating that the evolution is not fully adiabatic. However, despite the diabaticity with respect to the ground state, the evolution is still slow enough to enter the low-energy manifold of bound states. Extended Data Fig.\,\ref{fig:overlap}a characterizes these bound states, showing that the lowest 10 eigenstates are distinguished by an average hole-magnon distance of approximately one site. As shown in Extended Data Fig.\,\ref{fig:overlap}b, the calculated overlap of our final state with the subspace spanned by these 10 states is close to 1. This indicates that while the ramp is to fast to reach the ground-state, it is still able to prepare bound states with high fidelity.
\\
\\
\noindent{\bf{Negative temperature states.}} 
The initial product states we consider have large overlap with the lowest (highest) energy state in the presence of  strong light shifts with~$\delta_\sigma>0$ ($\delta_\sigma<0$). Therefore, by ramping down the magnitude of the light shifts, we quasi-adiabatically connect the system to the low-energy (high-energy) part of the many-body spectrum of~$\hat{H}_{tJ}$, preparing low (negative) temperature ensembles~\cite{Braun2013,CSB_Chen_2023,Annabelle_2024}.
An analogous perspective on the negative temperature states is obtained by considering the ground state of a Hamiltonian with reversed couplings, $\hat{H}_{tJ} \rightarrow -\hat{H}_{tJ}$. In the fermionic \tJ model with a single dopant, this sign flip corresponds to changing from hole to doublon doping.

To show that a model with a single {\it bosonic} hole dopant emulates the fermionic system with a hole or doublon dopant, depending on the sign of tunneling~$t$, we start by considering the fermionic \tJ model. First, we introduce parton operators, $\hat{c}_{i,\sigma}^\dagger = \hat{h}_i \hat{a}_{i,\sigma}^\dagger$, where the presence of a spinful fermion is associated with a bosonic spinon~$\hat{a}_{i,\sigma}^\dagger$ and the absence of a fermionic chargon~$\hat{h}_i$.
Then, the Hamiltonian of the fermionic \tJ model is given by (for simplicity with $J=0$)
\begin{equation}
\label{Fermi_Hubbard_H}
\begin{aligned}
&\hat{H}_{tJ}^{\text{fermion}} =-\hbar\sum_{i<j}\sum_{\sigma=\downarrow,\uparrow
}t_{\sigma}\frac{a^3}{{r}_{ij}^3}\hat{\mathcal{P}}_{G}\left(\hat{c}_{i,\sigma}^\dagger\hat{c}_{j,\sigma}+\text { h.c. }\right)\hat{\mathcal{P}}_{G}\\
&=+\hbar\sum_{i<j}\sum_{\sigma=\downarrow,\uparrow
}t_{\sigma}\frac{a^3}{{r}_{ij}^3}\hat{\mathcal{P}}_{G}\left(\hat{h}_{i}^\dagger \hat{h}_{j} \otimes \hat{a}^\dagger_{j,\sigma}\hat{a}_{i,\sigma}+\text { h.c. }\right)\hat{\mathcal{P}}_{G},
\end{aligned}
\end{equation}
where in the second line we have used the fermionic anti-commutation relation $\hat{h}_{i}^\dagger\hat{h}_{j} = -\hat{h}_{j}\hat{h}_{i}^\dagger$. For a single doublon dopant, described by the first line, the Hamiltonian is frustration free. Instead, for a single hole dopant, described by the second line, the Hamiltonian is kinetically frustrated on non-bipartite lattices. Thus for the {\it single} dopant case, where no additional fermionic exchange statistics changes matrix elements in the Hamiltonian, the two different cases are equivalent to the hard-core bosonic model by emulating the appropriate sign structure.

In conclusion, a bosonic \tJ model quantum simulator can realize a \textit{single} fermionic hole (doublon) dopant for $t>0$ ($t<0$). As described above, the ground states for the two different signs are obtained by preparing low (negative) temperature states that are characterized by kinetic antiferromagnetism (Nagaoka ferromagnetism).
\\
\\

\noindent{\bf State detection.}
Our readout method is based on mapping one or two of the three states on the presence of atoms in the final imaging of one experimental sequence (Fig. \ref{fig:setup}d(iii)). Given the interaction time scale of $2\pi/t\sim1\,\mu\text{s}$, we need first to stop the dynamics before the readout so there is no evolution during the measurement pulses. To do so, two freezing pulses are applied to send atoms in $\ket{h}$ and $\ket{\uparrow}$ to idle Rydberg manifolds, $\ket{G}\coloneq\ket{58G}$ and $\ket{P'}\coloneq\ket{61P}$ respectively, that do not interact with $\ket{\downarrow}$. Then by shining a $1014\,\si{\nano\meter}$ pulse  resonant on the transition $\ket{6P_{3/2}} \longleftrightarrow  \ket{\downarrow}$ during $6\,\mu\text{s}$, atoms in $\ket{\downarrow}$ will be deexcited toward the ground state manifold via the short-lived $\ket{6P_{3/2}}$ state. The tweezers are then turned back on trapping ground state atoms and expelling atoms left in the Rydberg manifold via the ponderomotive force. Thereby, atoms that were in the state $\ket{\downarrow}$ are imaged while the atoms that were in $\ket{h}$ or $\ket{\uparrow}$ are lost, allowing us to measure $\hat{n}^\downarrow $. This readout method is illustrated in Extended Data Fig.\,\ref{fig:sequences}a.

\begin{extfig*}
\mbox{}
\includegraphics[width=\textwidth]{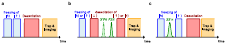}
\caption{\label{fig:sequences} 
{\bf Readout sequences.} 
\textbf{a,} Sequence to measure the $\ket{\downarrow}$ population, $\hat{n}^\downarrow$. The process begins with 'freezing' pulses that transfer the $\ket{h}$ and $\ket{\uparrow}$ states to the non-interacting $\ket{58G}$ and $\ket{61P}$ levels, respectively. A subsequent deexcitation pulse couples the remaining $\ket{\downarrow}$ atoms to the short-lived $\ket{6P}$ state, causing them to decay to the $\ket{5S}$ ground state. Finally, the optical tweezers are turned on, trapping the ground-state atoms for imaging while expelling all atoms that remain in Rydberg states. This procedure maps the $\ket{\downarrow}$ state onto the presence of an atom.
\textbf{b,} Similarly, to measure the population of $\ket{\uparrow}$ ($\hat{n}^\uparrow$) or $\ket{h}$ ($\hat{n}^h$), the sequence is modified to freeze the target state while deexciting the other two Rydberg states, thereby mapping the target state onto the absence of an atom (see text).
\textbf{c,} The sequence for the one-hole-two-magnon (1H2M) experiment also measures $\hat{n}^\uparrow$, but is designed to map the $\ket{\uparrow}$ state to the presence of an atom. This is achieved by first freezing the $\ket{h}$ state, then applying a microwave pulse to swap the $\ket{\uparrow}$ and $\ket{\downarrow}$ populations. Following the swap, the new $\ket{\uparrow}$ states (which were originally $\ket{\downarrow}$) are frozen. The deexcitation then acts on the remaining $\ket{\downarrow}$ states (which were originally $\ket{\uparrow}$), mapping them to the ground state for imaging.
}
\end{extfig*}

\begin{extfig*}
\mbox{}
\includegraphics[width=\textwidth]{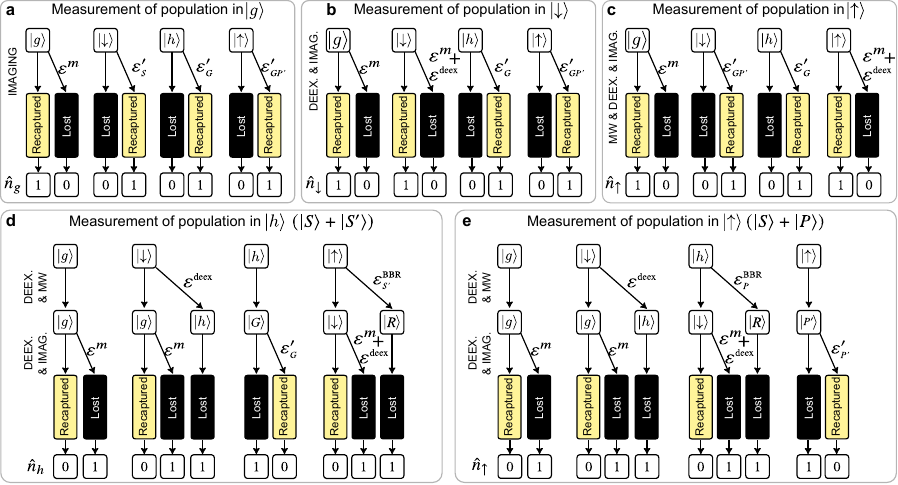}
\caption{\label{fig:error_tree} 
{\bf Error trees for the different readout methods.} The fives panels show the detection errors for the populations in $\ket{g}$, $\ket{\downarrow}$,$\ket{h}$ and $\ket{\uparrow}$. For each scheme we only show the detection events that affect the readout fidelity at first order. Meaning and value of the error channels are detailed in Tab.\ref{tab:error_values}. Panel \textbf{a,} corresponds to a simple measurement of the ground state population. The tweezers are directly turned on after the state preparation. \textbf{b,c,d,e}, Error trees corresponding to the readout sequences showed in Extended Fig.\ref{fig:sequences}a,c,b respectively. 
 }
\end{extfig*}

Because this final imaging gives us a binary result, atoms have either been recaptured or lost, we cannot distinguish the three states $\ket{\downarrow}$, $\ket{h}$ and $\ket{\uparrow}$ in a single shot. The readout method described above does not allow to distinguish between $\ket{h}$ and $\ket{\uparrow}$, thus we call it a $\ket{\downarrow}$-basis measurement. The $\ket{h}$- and $\ket{\uparrow}$-basis measurements are shown in Extended Data Fig.\,\ref{fig:sequences}(b,c). For the $\ket{h}$-basis measurement, after deexciting the $\ket{\downarrow}$ atoms, two $\pi$-pulses followed by a deexcitation pulse bring atoms in the state $\ket{\uparrow}$ toward the ground state. Atoms that were in $\ket{\downarrow}$ or $\ket{\uparrow}$ are imaged while the ones in $\ket{h}$ are lost. By considering lost atoms as hole, we measure the occupation of hole, $\hat{n}^{h}$, at each sites. The $\ket{\uparrow}$-basis measurement is realized by deexcitating atoms in $\ket{\downarrow}$, then bringing atoms in $\ket{h}$ to the state $\ket{\downarrow}$ via a microwave $\pi$-pulse and deexcite these atoms with a second pulse of the $1014\,\si{\nano\meter}$ laser. With this protocol, atoms that were in $\ket{\downarrow}$ or $\ket{h}$ are imaged while lost atoms can be interpret as atoms in $\ket{\uparrow}$ realizing a $\ket{\uparrow}$-basis measurement. It effectively implements the observable $\hat{n}^\uparrow $. In both cases a freezing microwave pulse is shone in order to send atoms in the state mapped onto an absence in the final imaging to another Rydberg manifold. It stops its interaction with the rest of the system and ensures that it will be expelled by the tweezer as it will remain in a Rydberg state.

To measure magnon-magnon correlations in the 1H2M case (Fig.\ref{fig:1H2M_bound_state}a) magnons have been directly mapped onto the presence of atoms following the sequence described in Extended Data Fig.\ref{fig:sequences}c. First atoms in the state $\ket{h}$ are sent to $\ket{G}$ and then a 2-photon process via the $\ket{60P}$ manifold realizes a $\pi$-pulse to invert $\ket{S}$ and $\ket{S'}$ propulation. Sending atoms in $\ket{S'}$ to $\ket{P'}$ thus effectively freezes atoms that were initially in $\ket{\downarrow}$. On the other hand, atoms that were in $\ket{\uparrow}$ are deexcited toward $\ket{g}$ to then be trapped and imaged realizing a measurement of $\hat{n}^\uparrow$.

In order to observe the antiferromagnetic order in the $x$-direction we must first rotate the spins before the measurement sequence. Therefore, before a $\ket{\downarrow}$-basis measurement a $\pi/2$-pulse between $\ket{\downarrow}$ and $\ket{\uparrow}$ is applied, which is realized by the 2-photon process. This allows us to implement the observable $\hat{n}^+$ defining $\ket{+}=(\ket{\uparrow}+\ket{\downarrow})/\sqrt{2}$. To measure the $\hat{n}^-$, we apply a $-\pi/2$-pulse between $\ket{\downarrow}$ and $\ket{\uparrow}$, which rotates the measurement basis by $\pi$ and thus corresponds to the measurement of $\hat{n}^-$. Then we can reconstruct $\hat{S}^{x}=(\hat{n}^{+}-\hat{n}^{-})/2$.
\\
\\

\noindent{\bf Error model.} 
To calibrate the initial state preparation errors, we perform measurement on this initial state in four basis: $\ket{\downarrow}$, $\ket{h}$, $\ket{\uparrow}$ and $\ket{g}$. The first three ones are realized as explained in the previous section (see also Extended Data Fig.\,\ref{fig:sequences}a,b) and the ground state basis measurement corresponds to a $\ket{\downarrow}$-basis measurement without deexcitation.

The errors results from finite vacuum, Rydberg lifetime, and imperfect detection fidelity (Extended Data Tab.\ref{tab:error_values}). To model them, we assume they are independent and we evaluate them at first order, leading to an error tree for each basis measurement, see Extended Data Fig.\,\ref{fig:error_tree}. Before the measurement sequence an atom is in one of the following five states: $\ket{\downarrow}$, $\ket{h}$, $\ket{\uparrow}$, $\ket{g}$ or $\ket{L}$ where $\ket{L}$ corresponds to a lost atom (due to a collision with the background gas for example). Given the distribution over these five states, the error trees give us the measured population for the four basis measurements:
\begin{equation*}
\label{eq:measured_pop}
\begin{aligned}
    \langle \hat{n}^\downarrow  \rangle^{\text{exp}} &= (1-\varepsilon^m)\mathcal{P}(g) + (1-\varepsilon^m-\varepsilon^{\text{deex}})\mathcal{P}(\downarrow) \\&+ \varepsilon'_G \mathcal{P}(P) + \varepsilon'_{GP'} \mathcal{P}(\uparrow) \\
    \langle \hat{n}^\uparrow  \rangle^{\text{exp}} &= \mathcal{P}(L) + \varepsilon^m \mathcal{P}(g) + (\varepsilon^m + \varepsilon^{\text{deex}})\mathcal{P}(\downarrow) \\&+ (\varepsilon^m + \varepsilon^{\text{deex}} + \varepsilon^{\text{BBR}}_P)\mathcal{P}(h) + (1-\varepsilon'_{P'})\mathcal{P}(\uparrow) \\
    \langle \hat{n}^h \rangle^{\text{exp}} &= \mathcal{P}(L) + \varepsilon^m \mathcal{P}(g) + (\varepsilon^m + \varepsilon^{\text{deex}})\mathcal{P}(\downarrow) \\&+ (1-\varepsilon'_G)\mathcal{P}(h) + (\varepsilon^m + \varepsilon^{\text{deex}} + \varepsilon^{\text{BBR}}_{\uparrow})\mathcal{P}(\uparrow) \\
    \langle \hat{n}^g \rangle^{\text{exp}} &= (1-\varepsilon^m)\mathcal{P}(g) + \varepsilon'_S \mathcal{P}(\downarrow) + \varepsilon'_G \mathcal{P}(h) + \varepsilon'_{GP'} \mathcal{P}(\uparrow)
\end{aligned}
\end{equation*}
where $\braket{\hat{n}_\alpha}^{\text{exp}}$ is the experimentally measured population in  state $\ket{\alpha}$ for a $\ket{\alpha}$-basis measurement, and $\mathcal{P}$ is the real probability of state. This yields  an error matrix:
\begin{equation}
\begin{bmatrix}
\langle \hat{n}^\downarrow  \rangle^{\text{exp}} \\
\langle \hat{n}^\uparrow  \rangle^{\text{exp}} \\
\langle \hat{n}^h \rangle^{\text{exp}} \\
\langle \hat{n}^g \rangle^{\text{exp}} \\
\end{bmatrix}
=
\varepsilon_{err}
\begin{bmatrix}
\mathcal{P}(L) \\
\mathcal{P}(g) \\
\mathcal{P}(S) \\
\mathcal{P}(P) \\
\mathcal{P}(S') \\
\end{bmatrix}
\end{equation}
with:
\begin{equation} \label{eq:error_matrix}
    \varepsilon_{err} = \begin{bmatrix}
0 & 0.998 & 0.978 & 0.013 & 0.02 \\
1 & 0.002 & 0.022 & 0.042 & 0.974 \\
1 & 0.002 & 0.022 & 0.987 & 0.042 \\
0 & 0.998 & 0.055 & 0.013 & 0.02 \\
\end{bmatrix}
\end{equation}
computed from the estimated error values shown in Extended Data Tab.~\ref{tab:error_values}.
\begin{table}[h!]
  \centering
  \begin{tabular}{|c|c|}
  \hline
  Error & Value \\
  \hline
  Mechanical losses + imaging fidelity & $\varepsilon^m=0.2\%$ \\
  0K lifetime of S state & $\varepsilon'_S = 5.5\%$ \\
  0K lifetime of the G state & $\varepsilon'_G=1.3\%$ \\
  0K lifetime of P' state & $\varepsilon'_{P'}=2.6\%$\\
  0K lifetime of a 50\% mixture of G and P' & $\varepsilon'_{GP'}=2\%$ \\
  Black body radiation for S' state & $\varepsilon^{\text{BBR}}_{S'}=2\%$ \\
  Black body radiation for P state & $\varepsilon^{\text{BBR}}_P = 2\%$ \\
  Deexcitation error & $\varepsilon^{\text{deex}}= 2\%$ \\
  \hline
  \end{tabular}
  \caption{\label{tab:error_values}{\bf Values of the errors used in the model.} Losses due to finite vacuum and imaging and the fidelity of the deexcitation are experementaly measured. Errors due to finite Rydberg lifetime are  computed numerically.
  }
\end{table}

\noindent{\bf Modeling the initial state.}
One challenge is to find an accurate description of the initial state prepared at time~$T=0\,\mu\mathrm{s}$ using a first principle model to numerically benchmark the experimental results without any fitting parameter. Moreover refined error models give insights into the experimental imperfections and further help improving state preparation and measurement (SPAM) errors.

In particular, the initial state prepared experimentally contains correlations, which are important to correctly describe beyond single-body observables; not including those correlated errors have been a limitation to our previous numerical model~\cite{tJ_qiao2025}. These correlated errors might be caused by, e.g. global power fluctuations or interactions between atoms during state preparation.
Experimentally, we can only measure in one basis state, making it theoretically difficult to reconstruct the initial state density matrix; this task is further complicated by the detection errors on the initial state. To this end, we develop a machine-learning inspired method that allows us to sample initial states, including correlations, from a Boltzmann machine.

We define an Ising-like energy functional (or Boltzmann machine) to a configuration~$s$, e.g. $s=\ket{...\downarrow \uparrow g \uparrow L h \downarrow ... }$, given by
\begin{equation}
    \label{boltzman_machine}
    E(s)=-\sum_{\substack{i<j \\ i,j \in \Omega}}\sum_{\alpha}J^{\alpha}_{ij} n^{\alpha}_i n^{\alpha}_j - \sum_j \sum_{\alpha} h^\alpha_j n^{\alpha}_j ,
\end{equation}
where the region~$\Omega$ is a subset of sites around the initial hole and magnon site. The index $\alpha$ denotes the fives internal states of a sites, including the three physical states of the $t$-$J$~model as well as atoms in the atomic ground state~$\ket{g}$ and lost atoms~$\ket{L}$. We only consider configurations that are diagonal in the computational basis.

For a given set of parameters~$K = \{ J_{ij}, h_j \}$, we compute the partition function
\begin{align}
    Z = \sum_{s} e^{-E(s)}
\end{align}
and statistical probability~$p(s) = e^{-E(s)}/Z$ for a configuration~$s$.
Hence, we can determine all (thermal) correlation functions of the model, e.g. the two-point correlator
\begin{align}
    \langle \hat{n}^\alpha_i \hat{n}^\alpha_j \rangle = \sum_{s} \sum_{\beta}([\varepsilon_{err}]_\beta^\alpha\delta_{s_i,n^\beta_i}) ([\varepsilon_{err}]_\beta^\alpha\delta_{s_j,n^\beta_j}) \frac{e^{-E(s)}}{Z}.
\end{align}
When evaluating the expectation values, we have included the error matrix~$\varepsilon_{err}$, see Eq.~\eqref{eq:error_matrix}, that mixes the measurements in a non-unitary way, and encodes the read-out and detections errors discussed above.

Our goal is to find the parameters~$K$ using a maximum entropy principle. 
We define a cost function with the following contributions:
\begin{align}
    \epsilon^{\rm one} &= \sum_j \sum_\alpha | \langle \hat{n}^\alpha_j\rangle_{\rm exp} - \langle \hat{n}^\alpha_j\rangle_{\rm fit} |^2 \\
    \epsilon^{\rm two} &= \sum_{\substack{i<j \\ i,j \in \Omega}}\sum_{\alpha} | \langle \hat{n}^\alpha_i \hat{n}^\alpha_j\rangle_{\rm exp} - \langle \hat{n}^\alpha_i \hat{n}^\alpha_j\rangle_{\rm fit} |^2 \\
    \epsilon^{\rm lost} &= \sum_j \langle \hat{n}^L_j\rangle_{\rm fit} \\
    \epsilon^{\rm target} &= 1 - p(s^{\rm target}).
\end{align}
In particular, we consider two cost functions given by a weighted sum of the above contributions
\begin{align}
    \mathrm{cost}^{\rm fit} &= w^{\rm one}\epsilon^{\rm one} + w^{\rm two}\epsilon^{\rm two} + w^{\rm lost}\epsilon^{\rm lost} + w^{\rm target}\epsilon^{\rm target} \\
    \mathrm{cost}^{\rm true} &= \epsilon^{\rm one} + \epsilon^{\rm two}.
\end{align}
The minimization algorithm aims at finding the global minimum of the cost function~$\mathrm{cost}^{\rm true}$. Since the problem is high dimensional, given the dimension of the configuration space and the number of parameters scaling as $\mathcal{O}(|\Omega|^2)$, we can only find a good estimate. Therefore, we iterate different weights in~$\mathrm{cost}^{\rm fit}$ to help the system finding a minimum that shows good agreement with the experimental data. We notice that in our minimization procedure we were only able to find local minima that either minimize the correlation functions, with the cost of maximizing the single particle errors, or vice versa. 

\begin{extfig}
\mbox{}
\includegraphics[width=\linewidth]{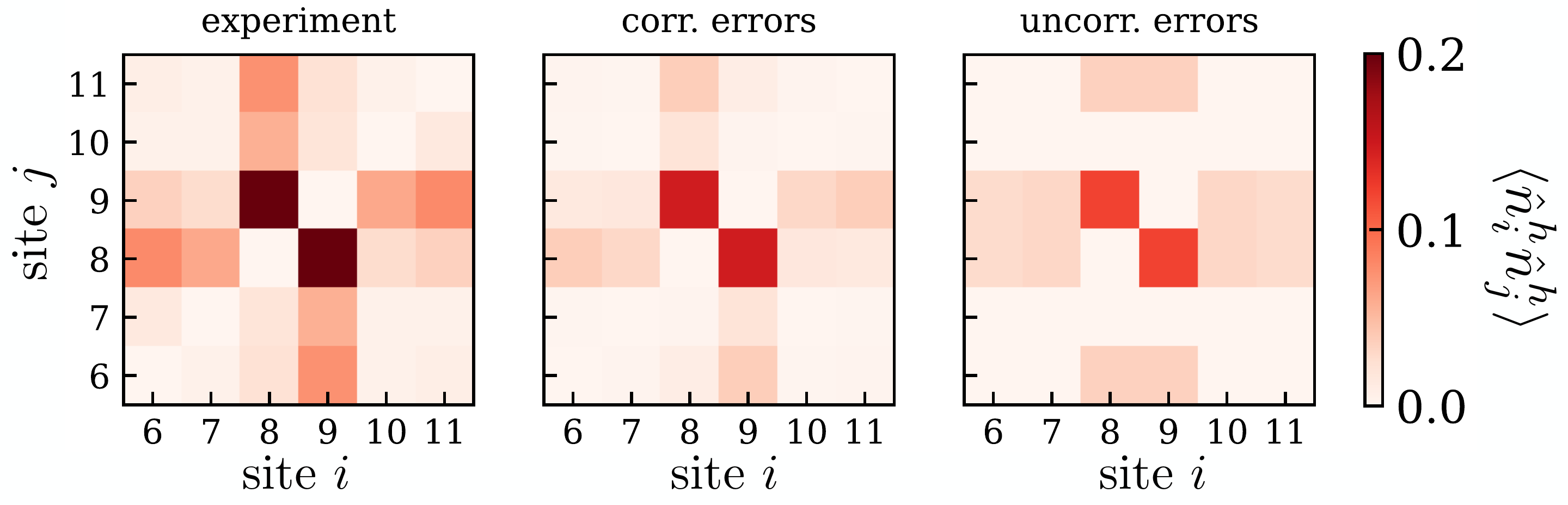}
\caption{\label{fig:corr_errors} 
{\bf Correlated errors.} We plot the hole-hole correlations in a region of interest~$\Omega$. The experimental measurements at time~$T=0\,\mu\mathrm{s}$ (left) show strong correlations of two neighboring holes, and additional faint structure in the correlation map. Our model based on a Boltzmann machine (middle) is able to qualitatively capture those correlations, improving our understanding of the experimental limitations. The simplest model of an uncorrelated, single site error budget (right) only describes the main features in the correlations. }
\end{extfig}

We apply the Boltzmann machine approach to the numerically tractable cases of 1H1M and 1H2M in the triangular ladder. To performance of finding the minimum of the cost function depends on the dimensionality of the Hilbert space. Since we do not expect any events where, e.g. all atoms are falsely prepared in the hole state, we restrict the underlying Hilbert space~$\{ s \}$: For atoms in the outer regions, i.e. ~$j \notin \Omega$, the energy functional is independent of the rest and hence we include all five local states in the computation. For atoms in the correlated region, i.e. ~$j \in \Omega$, we consider a restricted Hilbert space, including configuration with total number of holes within $0 \leq N^h \leq 3$, a total spin within $-|\Omega|/2 \leq S^z_{\rm tot} \leq 1/2$, and total number of ground state atoms within $0 \leq N^{gs} \leq 2$.

In Fig.~\ref{fig:corr_errors}, we show an example of non-connected hole-hole correlations~$\langle \hat{n}^h_i \hat{n}^h_j \rangle$ within the region~$\Omega \in [6,11]$, for the 1H1M experiment. The target initial state has a magnon at site~$8$ and a hole at site~$9$. We observe a strong correlation to falsely prepare two neighboring holes, and smaller but visible correlation to prepare two holes next-nearest neighbor sites. Extracting an error budget for each site individually, i.e. when $J_{ij} \equiv 0$ in Eq.~\eqref{boltzman_machine}, already captures substantial correlations in the initial state. However, the subleading structure in the correlation map -- important to quantitatively describe the dynamics in the experiment -- cannot be accounted for. In contrast, the correlated error map does to some extent correctly include these correlations. 

Currently, the ability to perform state tomography on the initial state is limited by detection errors and basis measurements in only one basis. While this approach improves our understanding of the limitations in our experiment, it further opens new directions to study the thermodynamics of our experiments~\cite{Sbierski2024}. For instance, a more elaborate model including higher-order correlators could allow one to extract the entropy of the initial state, and analyze how close the initial state is to a thermal distribution.
\\
\\

\noindent{\bf Correlation reconstruction.} 
As we cannot distinguish between the three states $\ket{\uparrow}$, $\ket{h}$ and $\ket{\downarrow}$ in a single shot, 
we reconstruct the two-body and three-body correlations using the procedure detailed below. In an ideal system, each site can be in three possible states, leading to a total of nine different two-body density correlators between sites $i$ and $j$: $\braket{\nhat_{\sigma} \nhat_{\tau}}$, where $\sigma, \tau \in \{\uparrow, \downarrow, h\}$. Experimentally, we perform projective measurements in three distinct bases: the $\ket{\uparrow}$, the $\ket{\downarrow}$, and the $\ket{h}$-basis, as described before.

For each basis $\sigma \in \{\uparrow, \downarrow, h\}$, a single shot measurement can only give outcomes $\sigma$ or $\left.\not{\sigma}\right.$ (not $\sigma$) at each site. Averaging over many shots we could get twelve distinct two-site measurement outcomes: $\braket{\sigma \sigma}$, $\braket{\sigma \not\sigma}$, $\braket{\not\sigma \sigma}$, and $\braket{\not\sigma \not\sigma}$ for each of the three bases. These measured probabilities can be expressed as linear combinations of the 2-body correlators $\braket{\nhat_{\sigma} \nhat_{\tau}}$. Using the completeness relation $\nhat_{\uparrow,k} + \nhat_{\downarrow,k} + \nhat_{h,k} = \hat{\mathbb{I}}_k$ for any site $k$, we have:

\begin{align*}
&\braket{\uparrow\uparrow}=\braket{\hat{n}^{\uparrow}\hat{n}^{\uparrow}}\\
&\braket{\uparrow{\not{\uparrow}}}=\braket{\hat{n}^{\uparrow}\hat{n}^{h}}+\braket{\hat{n}^{\uparrow}\hat{n}^{\downarrow}}\\
&\braket{{\not{\uparrow}}\uparrow}=\braket{\hat{n}^{h}\hat{n}^{\uparrow}}+\braket{\hat{n}^{\downarrow}\hat{n}^{\uparrow}}\\
&\braket{{\not{\uparrow}}{\not{\uparrow}}}=\braket{\hat{n}^{h}\hat{n}^{h}}+\braket{\hat{n}^{h}\hat{n}^{\downarrow}}+\braket{\hat{n}^{\downarrow}\hat{n}^{h}}+\braket{\hat{n}^{\downarrow}\hat{n}^{\downarrow}}\\
&\braket{hh}=\braket{\hat{n}^{h}\hat{n}^{h}}\\
&\braket{h{\not{h}}}=\braket{\hat{n}^{h}\hat{n}^{\uparrow}}+\braket{\hat{n}^{h}\hat{n}^{\downarrow}}\\
&\braket{{\not{h}}h}=\braket{\hat{n}^{\uparrow}\hat{n}^{h}}+\braket{\hat{n}^{\downarrow}\hat{n}^{h}}\\
&\braket{{\not{h}}{\not{h}}}=\braket{\hat{n}^{\uparrow}\hat{n}^{\uparrow}}+\braket{\hat{n}^{\uparrow}\hat{n}^{\downarrow}}+\braket{\hat{n}^{\downarrow}\hat{n}^{\uparrow}}+\braket{\hat{n}^{\downarrow}\hat{n}^{\downarrow}}\\
&\braket{\downarrow\downarrow}=\braket{\hat{n}^{\downarrow}\hat{n}^{\downarrow}}\\
&\braket{\downarrow{\not{\downarrow}}}=\braket{\hat{n}^{\downarrow}\hat{n}^{h}}+\braket{\hat{n}^{\downarrow}\hat{n}^{\uparrow}}\\
&\braket{{\not{\downarrow}}\downarrow}=\braket{\hat{n}^{\uparrow}\hat{n}^{\downarrow}}+\braket{\hat{n}^{h}\hat{n}^{\downarrow}}\\
&\braket{{\not{\downarrow}}{\not{\downarrow}}}=\braket{\hat{n}^{\uparrow}\hat{n}^{\uparrow}}+\braket{\hat{n}^{\uparrow}\hat{n}^{h}}+\braket{\hat{n}^{h}\hat{n}^{\uparrow}}+\braket{\hat{n}^{h}\hat{n}^{h}}
\end{align*}

These relations can be written in matrix form $\vec{M} = \mathbf{A} \vec{C}$, where $\vec{M}$ is the column vector of the 12 measurable quantities (e.g., $[\braket{\uparrow \uparrow }, \braket{\uparrow {\not{\uparrow }}}, \dots, \braket{{\not{\downarrow }}{\not{\downarrow }}}]^T$) and $\vec{C}$ is the column vector of the nine 2-body correlators (e.g., $[\braket{\nhat_\uparrow \nhat_\uparrow}, \braket{\nhat_\uparrow \nhat_h}, \dots, \braket{\nhat_\downarrow \nhat_\downarrow}]^T$). The $12 \times 9$ matrix $\mathbf{A}$ is given by:

\begin{extfig*}
\mbox{}
\includegraphics[width=0.9\textwidth]{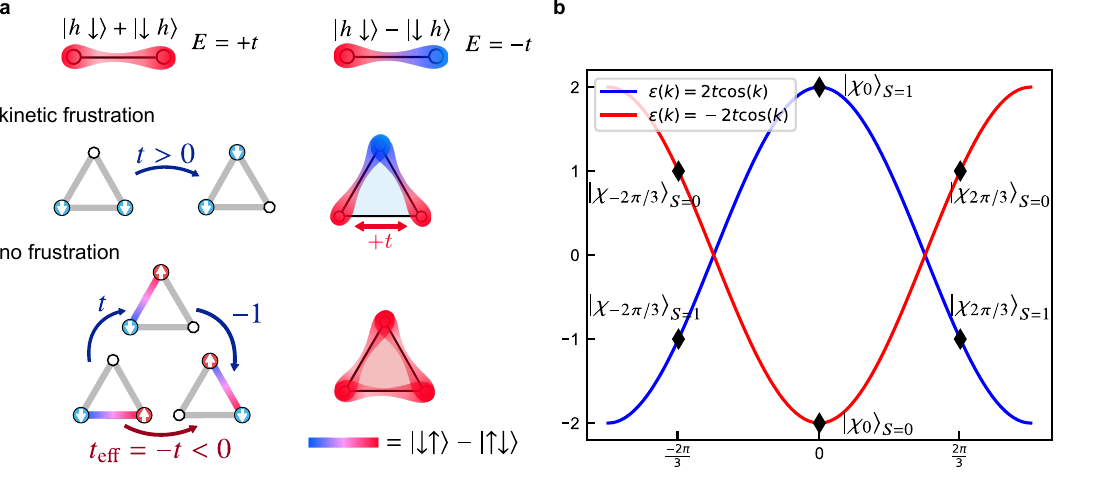}
\caption{\label{fig:toy_model} 
{\bf Toy model on a triangular plaquette. a,} For two sites, the symmetric wavefunction has an energy of $E=+t$, and the antisymmetric wavefunction has an energy of $E=-t$. On a triangle, a single hole with positive tunneling ($t>0$) experiences kinetic frustration, which raises its kinetic energy. A nearby magnon forms a singlet state with an adjacent spin, effectively reversing the tunneling sign ($t^{\text{eff}} \approx -t < 0$). This relieves the frustration, lowering the kinetic energy. \textbf{b, } Eigenstates and band structures of one fermionic hole hopping on 3 sites in a background forming a spin triplet (blue) or spin singlet (red).}
\end{extfig*}

\begin{equation}
\label{eq:two-body correlator}
\left(\begin{array}{c}\braket{\uparrow \uparrow }\\ \braket{\uparrow {\not{\uparrow }}}\\ \braket{{\not{\uparrow }}\uparrow }\\ \braket{{\not{\uparrow }}{\not{\uparrow }}}\\ \braket{hh}\\ \braket{h{\not{h}}}\\ \braket{{\not{h}}h}\\ \braket{{\not{h}}{\not{h}}}\\ \braket{\downarrow \downarrow }\\ \braket{\downarrow {\not{\downarrow }}}\\ \braket{{\not{\downarrow }}\downarrow }\\ \braket{{\not{\downarrow }}{\not{\downarrow }}} \end{array}\right)=\left(\begin{array}{lllllllll}1 & 0 & 0 & 0 & 0 & 0 & 0 & 0 & 0 \\ 0 & 1 & 1 & 0 & 0 & 0 & 0 & 0 & 0 \\ 0 & 0 & 0 & 1 & 0 & 0 & 1 & 0 & 0 \\ 0 & 0 & 0 & 0 & 1 & 1 & 0 & 1 & 1 \\ 0 & 0 & 0 & 0 & 1 & 0 & 0 & 0 & 0 \\ 0 & 0 & 0 & 1 & 0 & 1 & 0 & 0 & 0 \\ 0 & 1 & 0 & 0 & 0 & 0 & 0 & 1 & 0 \\ 1 & 0 & 1 & 0 & 0 & 0 & 1 & 0 & 1 \\ 0 & 0 & 0 & 0 & 0 & 0 & 0 & 0 & 1 \\ 0 & 0 & 0 & 0 & 0 & 0 & 1 & 1 & 0 \\ 0 & 0 & 1 & 0 & 0 & 1 & 0 & 0 & 0 \\ 1 & 1 & 0 & 1 & 1 & 0 & 0 & 0 & 0\end{array}\right)\left(\begin{array}{l}\left\langle \hat n_\uparrow  \hat n_\uparrow \right\rangle \\ \left\langle \hat n_\uparrow  \hat n_h\right\rangle \\ \left\langle \hat n_\uparrow  \hat n_\downarrow \right\rangle \\ \left\langle \hat n_h \hat n_\uparrow \right\rangle \\ \left\langle \hat n_h \hat n_h\right\rangle \\ \left\langle \hat n_h \hat n_\downarrow \right\rangle \\ \left\langle \hat n_\downarrow  \hat n_\uparrow \right\rangle \\ \left\langle \hat n_\downarrow  \hat n_h\right\rangle \\ \left\langle \hat n_\downarrow  \hat n_\downarrow \right\rangle\end{array}\right)
\end{equation}

The rank of this matrix $\mathbf{A}$ is 8 and the correlation vector $\vec{C}$ is 9-dimension, therefore our measurements are insufficient to uniquely determine all nine correlators. The measured data $\vec{M}$ constrain the correlators $\vec{C}$ to an 8-dimensional subspace, leaving an ambiguity corresponding to the null space of $\mathbf{A}$. To fully reconstruct $\vec{C}$, we would need additional cross-basis measurements. However, certain linear combinations of the 2-body correlators can still be determined solely from our measurements by solving the linear equation Eq.\,\ref{eq:two-body correlator}:

\begin{itemize}
    \item Spin-spin correlation: Defining $\hat{S}_{z,k} = (\hat{n}_{\uparrow,k} - \hat{n}_{\downarrow,k})/2$, the correlator is $4\braket{\hat{S}^{z}\hat{S}^{z}} = \braket{\nhat^{\uparrow}\nhat^{\uparrow}} - \braket{\nhat^{\uparrow}\nhat^{\downarrow}} - \braket{\nhat^{\downarrow}\nhat^{\uparrow}} + \braket{\nhat^{\downarrow}\nhat^{\downarrow}}= \braket{\uparrow\uparrow} + \braket{\not{\uparrow}\not{\uparrow}} - \braket{{h}{h}} - \braket{\downarrow\not{\downarrow}} - \braket{\not{\downarrow}\downarrow}$.

    \item Symmetric hole-spin correlation: $2(\braket{\hat{n}^{h}\hat{S}^{z}} + \braket{\hat{S}^{z}\hat{n}^{h}}) = \left(\braket{\nhat^{h}\nhat^{\uparrow}} - \braket{\nhat^{h}\nhat^{\downarrow}} + \braket{\nhat^{\uparrow}\nhat^{h}} - \braket{\nhat^{\downarrow}\nhat^{h}}\right)=\left(-\braket{\uparrow {\uparrow}} - \braket{{\not{\uparrow }}\not{\uparrow}} + \braket{\downarrow {\downarrow }} + \braket{{\not{\downarrow }}\not{\downarrow} }\right)$.

    \item Symmetric hole-up correlation: $\braket{\hat{n}^{h}\hat{n}^{\uparrow}} + \braket{\hat{n}^{\uparrow}\hat{n}^{h}}=\braket{{\not{\downarrow} {\not{\downarrow }}}} - \braket{\uparrow \uparrow } - \braket{hh}$.
\end{itemize}

The method for reconstructing two-body correlators can be extended to 3-body correlators where we have $27$ different 3-body correlators $\braket{\nhat_{\sigma_1} \nhat_{\sigma_2} \nhat_{\sigma_3}}$ but only $24$ possible measurements. However, similar to the 2-body case, specific linear combinations of the 3-body correlators can be reconstructed, such 
as the Symmetric hole-spin-spin correlator:
     $4\left(\braket{\hat{n}^{h}\hat{S}^{z}\hat{S}^{z}} + \braket{\hat{S}^{z}\hat{n}^{h}\hat{S}^{z}} + \braket{\hat{S}^{z}\hat{S}^{z}\hat{n}^{h}}\right) = \braket{\hat{n}^{h}\hat{n}^{\uparrow}\hat{n}^{\uparrow}} - \braket{\hat{n}^{h}\hat{n}^{\uparrow}\hat{n}^{\downarrow}} - \braket{\hat{n}^{h}\hat{n}^{\downarrow}\hat{n}^{\uparrow}} + \braket{\hat{n}^{h}\hat{n}^{\downarrow}\hat{n}^{\downarrow}} + \braket{\hat{n}^{\uparrow}\hat{n}^{h}\hat{n}^{\uparrow}} - \braket{\hat{n}^{\uparrow}\hat{n}^{h}\hat{n}^{\downarrow}} - \braket{\hat{n}^{\downarrow}\hat{n}^{h}\hat{n}^{\uparrow}} + \braket{\hat{n}^{\downarrow}\hat{n}^{h}\hat{n}^{\downarrow}} + \braket{\hat{n}^{\uparrow}\hat{n}^{\uparrow}\hat{n}^{h}} - \braket{\hat{n}^{\uparrow}\hat{n}^{\downarrow}\hat{n}^{h}} - \braket{\hat{n}^{\downarrow}\hat{n}^{\uparrow}\hat{n}^{h}} + \braket{\hat{n}^{\downarrow}\hat{n}^{\downarrow}\hat{n}^{h}}=\braket{\uparrow\uparrow\!\not{\uparrow}} + \braket{\uparrow\!\not{\uparrow}\uparrow} + \braket{\!\not{\uparrow}\uparrow\uparrow} + \braket{\!\not{\uparrow}\!\not{\uparrow}\!\not{\uparrow}} - \braket{hhh} - \braket{\downarrow\downarrow\downarrow} - \braket{\downarrow\!\not{\downarrow}\!\not{\downarrow}} - \braket{\!\not{\downarrow}\downarrow\!\not{\downarrow}} - \braket{\!\not{\downarrow}\!\not{\downarrow}\downarrow}$.
\\
\\

\noindent{\bf Toy model on a triangular plaquette.} To understand the principle of the kinetically-induced binding, we consider a triangular plaquette in two different configurations, illustrated in Fig.~\ref{fig:toy_model}a: one hole and a background of two spins down; or one hole with a background composed of one magnon and one spin down. We furthermore neglect the interaction ($J_\perp=J_z=0$) between spins and only keep the hopping part~$\propto t$ of the Hamiltonian. 

In the polarized spin background, the Hamiltonian then reduces to the one of a free particle on a 3-sites triangular plaquette with eigenstates~\cite{OL_Lebrat2024}:

\begin{equation}
\ket{\chi_k} = \left( \ket{h\downarrow\downarrow} + e^{ik} \ket{\downarrow h\downarrow} + e^{i2k}\ket{\downarrow\downarrow h}\right)/\sqrt{3}
\end{equation}
with momentum $k \in \{0; 2\pi/3;-2\pi/3 \}$, and corresponding energies $\varepsilon(k)=2t\cos k$. The states and the band structure are shown in Extended Data Fig.\,\ref{fig:toy_model}. The frustration arises from the effective sign of the tunneling $+t$ of the hole.  Crucially,  noting that the two spins form a triplet $\ket{T}=\ket{\downarrow \downarrow}$, one can rewrite these states as follows:

\begin{equation}
\ket{\chi_k} = \left(  \ket{h}_1\ket{T}_{23} + e^{ik}\ket{h}_2\ket{T}_{31} + e^{i2k}\ket{h}_3\ket{T}_{12} \right)/ \sqrt{3}
\end{equation}
where $\ket{T}_{ij}$ is a triplet state between sites $i$ and $j$.

Turning now to the case of one hole, one magnon and one spin down, the give rise to an additional degree-of-freedom in the wavefunction. Since the Hamiltonian conserves the total spin $S^2 = S_x^2+S_y^2+S_z^2$, we can label the eigenstates by singlet states ($S =0$) and triplet states ($S=1$). The singlet subspace can be written as: $\ket{h}_1\ket{s}_{23}; \ket{h}_2\ket{s}_{31}; \ket{h}_3\ket{s}_{12}$ with $\ket{s}_{ij}= \left( \ket{\downarrow_i\uparrow_j} - \ket{\uparrow_i \downarrow_j} \right) / \sqrt{2}$. Similarly, the triplet subspace $S=1$ is given by: $\ket{h}_1\ket{T_0}_{23}; \ket{T_0}_2\ket{s}_{31}; \ket{h}_3\ket{T_0}_{12}$ with $\ket{T_0}_{ij}= \left( \ket{\downarrow_i\uparrow_j} + \ket{\uparrow_i \downarrow_j} \right) / \sqrt{2}$. From this, we can compute the effective tunneling amplitudes for the hole by considering the following matrix elements:
\begin{equation}
    \begin{aligned}
        &\bra{h}_i\bra{T_0}_{jk}\hat{H}_t\ket{h}_{i'}\ket{T_0}_{j'k'}=t \\
        &\bra{h}_i\bra{s}_{jk}\hat{H}_t\ket{h}_{i'}\ket{s}_{j'k'}=-t
    \end{aligned}
\end{equation}
where $(i,j,k)$ and $(i',j',k')$ are two different permutations of $(1,2,3)$. The additional negative sign in the second equation originates from the antisymmetric singlet wavefunction under permutation. Therefore, in the triplet subspace one retrieves the same sign structure of tunnelings as in the case without magnon, while in the singlet subspace we obtain frustration-free tunnelings: The singlet state effectively reverses the band structure in the case of a single triangular plaquette. 

The eigenstates in the singlet and triplet subspace are then given by
\begin{equation}
\begin{aligned}
&\ket{\chi_k}_{S=1} = (  \ket{h}_1\ket{T_0}_{23} + e^{ik}\ket{h}_2\ket{T_0}_{31} + e^{i2k}\ket{h}_3\ket{T_0}_{12}\\
&\ket{\chi_k}_{S=0} = (  \ket{h}_1\ket{s}_{23} + e^{ik}\ket{h}_2\ket{s}_{31} + e^{i2k}\ket{h}_3\ket{s}_{12}
\end{aligned}
\end{equation}
with energies $\varepsilon_{S=1}(k)=2t\cos k$ for the triplet states and $\varepsilon_{S=0}(k)=-2t\cos k$ for the singlet states. The eigenenergies are shown in Fig.~\ref{fig:toy_model}b. Kinetic frustration is relieved by the possibility to form singlets around the hole which effectively reverses the sign of the hopping term and allows the hole to retrieve its full kinetic energy of $-2t$ in this toy model. The presence of the magnon next to the hole is thus energetically favorable and explains the binding between the two particles, which we analytically showed for a single triangle. Furthermore, the spins forming a singlet in the ground state gives an intuition for our observation of anti-ferromagnetism in both the $x$ and $z$ directions around the hole.
\\
\\

\subsection*{Acknowledgements}
We thank Guillaume Bornet, Cheng Chen, and Gabriel Emperauger for early experimental contributions; Ana Maria Rey and Adam Kaufman for fruitful discussions; and Andrea Fantini for help with softwares.
This work is supported by
the Agence Nationale de la Recherche (ANR-22-PETQ-0004 France 2030, project QuBitAF), 
the European Research Council (Advanced grant No. 101018511-ATARAXIA), and 
the Horizon Europe programme HORIZON-CL4- 2022-QUANTUM-02-SGA (project 101113690 (PASQuanS2.1). R.M. acknowledges support by the ‘Fondation CFM pour la Recherche’ through a Jean-Pierre Aguilar PhD scholarship. L.H. acknowledges support by the Simons Collaboration on Ultra-Quantum Matter, which is a grant from the Simons Foundation (651440). I.M. and E.D. were supported by the SNSF (project 200021\_212899), the Swiss State Secretariat for Education, Research and Innovation (contract number UeM019-1) and the ARO (grant no. W911NF-20-1-0163).

\section*{Data Availability}

All data are available from the corresponding author on request.

\section*{Competing interests}
A.B. and T.L. are cofounders and shareholders of PASQAL. The remaining authors declare no competing interests.

\section*{Author contributions}
M.Q., R.M., L.H. and I.M. contributed equally to this work.
M.Q. and R.M. designed and carried out the experiments, with the help
of B.G., L.K., C.T., and D.B. L.H. and I.M. conducted the conceptual and theoretical analysis, and the numerical simulations. 
T.L., E.D. and A. B. supervised the work. All authors
contributed to the data analysis, progression of the project and on both the experimental and
theoretical sides. All authors contributed to the writing of the manuscript. 
Correspondence
and requests for materials should be addressed to Antoine Browaeys or Mu Qiao.

\end{document}